\documentclass{class}
\usepackage{graphicx}
\usepackage{epstopdf, epsfig}
\usepackage{siunitx}
\usepackage{amssymb} 
\usepackage{dirtytalk}
\usepackage{textcomp}
\newcommand{\solidline}{\rule[0.5ex]{.5cm}{0.5pt}}
\newcommand{\dashedline}{\rule[0.5ex]{.1cm}{0.5pt}\hspace{0.1cm}\rule[0.5ex]{.1cm}{0.5pt}\hspace{0.1cm}\rule[0.5ex]{.1cm}{0.5pt}}
\newcommand{\solidlinediamond}{{\scriptsize\rule[0.5ex]{.15cm}{0.5pt}}{\scriptsize$\blacklozenge$}{\scriptsize\rule[0.5ex]{.15cm}{0.5pt}}}
\newcommand{\solidlinesquare}{\rule[0.5ex]{.15cm}{0.5pt}{\tiny$\blacksquare$}\rule[0.5ex]{.15cm}{0.5pt}}
\newcommand{\solidlinecirc}{{\tiny\rule[0.5ex]{.15cm}{0.5pt}}{\tiny$\ominus$}{\tiny\rule[0.5ex]{.15cm}{0.5pt}}}
\newcommand{\squaresymb}{\hspace{0.1cm}{\tiny$\blacksquare$}\hspace{0.1cm}}

\newcommand{\circlesymb}{\hspace{0.1cm}{\large$\bullet$}\hspace{0.1cm}}
\newcommand{\solidlinesolidcirc}{{\rule[0.5ex]{.15cm}{0.5pt}}{\large$\bullet$}{\rule[0.5ex]{.15cm}{0.5pt}}}
\newcommand{\dashedlinecircle}{{\tiny\rule[0.5ex]{.1cm}{0.5pt}}\hspace{0.1cm}{\tiny\rule[0.5ex]{.1cm}{0.5pt}}\hspace{0.1cm}{\tiny\rule[0.5ex]{.1cm}{0.5pt}}\hspace{0.1cm}\hspace{-0.45cm}{\tiny$\ominus$}\hspace{0.15cm}}
\newcommand{\solidtriangle}{{\scriptsize$\blacktriangle$}}
\newcommand{\solidlefttriangle}{{\scriptsize$\blacktriangleright$}}
\newcommand{\hollowdiamond}{{\scriptsize$\lozenge$}}
\newcommand{\hollowcircle}{{\large\textopenbullet}}

\title{On the Low-Frequency Dynamics of Turbulent Separation Bubbles}
\author{C. Cura\aff{1},
  A. Hanifi\aff{2},   A. V. G. Cavalieri\aff{3}
 \and J. Weiss\aff{1}}

\affiliation{\aff{1}Insitut für Luft- und Raumfahrt, Technische Universit\"at Berlin, 10587 Berlin, Germany
\aff{2}FLOW, Engineering Mechanics, KTH Royal Institute of Technology, 100 44 Stockholm, Sweden
\aff{3} Divis\~ao de Engenharia Aeronáutica, Instituto Tecnológico de Aeronáutica, 12228-900 S\~ao José dos Campos, SP, Brazil}
\begin{document}
\maketitle
\section*{Abstract}
\begin{abstract}
The low-frequency modal and non-modal stability characteristics of an incompressible, pressure-gradient-induced turbulent separation bubble (TSB) are investigated with the objective of studying the mechanism responsible for the low-frequency contraction and expansion (breathing) commonly observed in experimental studies. The configuration of interest is a TSB generated on a flat test surface by a succession of adverse and favourable pressure gradients. The base flow selected for the analysis is the average TSB from the direct numerical simulation of Coleman \textit{et al.} (\textit{J. Fluid Mech.}, vol. 847, 2018). Global linear stability analysis reveals that the flow is globally stable for all frequencies and wavenumbers. Furthermore, the mode closest to the stability threshold appears to occur at zero frequency and low, non-zero spanwise wavenumber when scaled with the separation length.  Resolvent analysis is then employed to examine the forced dynamics of the flow. At low frequency, a region of low, non-zero spanwise wavenumber is also discernible, where the receptivity appears to be driven by the identified weakly damped global mode. The corresponding optimal energy gain is shown to have the shape of a first-order, low-pass filter with a cut-off frequency consistent with the low-frequency unsteadiness in TSBs. The results from resolvent analysis are compared to the unsteady experimental database of Le Floc'h \textit{et al.} (\textit{J. Fluid Mech.}, vol. 902, 2020) in a similar TSB flow. The alignment between the optimal response and the first spectral proper orthogonal decomposition mode computed from the experiments is shown to exceed $\SI{95}{\percent}$, while the spanwise wavenumber of the optimal response is consistent with that of the low-frequency breathing motion captured experimentally. This indicates that the fluctuations observed experimentally at low frequency closely match the response computed from resolvent analysis. Based on these results, we propose that the forced dynamics of the flow, driven by the weakly damped global mode, serve as a plausible mechanism for the origin of the low-frequency breathing motion commonly observed in experimental studies of TSBs.  
\end{abstract}
\section{Introduction}

Flow separation is a common phenomenon within fluid dynamics, which arises when a fluid flow is no longer able to follow the trajectory imposed by a solid wall. Such flows exhibit a variety of detrimental effects, including reduced lift, increased drag, noise emission or vibrations, all of which may negatively impact the performance of the system under consideration. A distinct subset within separated flows is the category of reattaching flows, where the flow reattaches again to the wall, e.g., due to surface curvature or a favorable pressure gradient. This category gives rise to so-called separation bubbles, which are known to feature unsteadiness in a variety of spatial and temporal scales. In the present contribution we specifically consider turbulent separation bubbles (TSB), which occur when a turbulent boundary layer separates from the wall and reattaches further downstream. We further focus our study on pressure-gradient-induced TSBs, where detachment from a smooth surface occurs because of an adverse pressure gradient \citep{namoin}. This is in contrast to flows where the separation line is fixed by the surface geometry \citep{eaton1981}.

Unsteadiness in pressure-induced TSBs typically occurs in three broad ranges of frequencies that may be categorized by their Strouhal number based on the separation length $L_b$ and a reference velocity $U_{\rm ref}$ \citep{Mabey1972}. For relatively high values $St > 1$, fluctuations are mainly caused by turbulent motions that have their highest amplitude in the attached flow upstream and downstream of the backflow region \citep{abe2017,le2020, wu2020}. Within the recirculation zone, medium frequencies centered at $St \simeq 0.1-1.0$ appear in the wall-pressure and velocity fields due to the roll-up and shedding of vortices originating in the shear layer \citep{kiya,cherry}. Finally, a low-frequency unsteadiness, typically characterized by a large-scale contraction and extension (\say{breathing}) of the TSB, is often observed at $St \simeq 0.01-0.1$ \citep{mohammed2016}. This is the main focus of the present work.

To date, low-frequency unsteadiness in pressure-induced TSBs has mostly been observed in high-speed flows, where it often occurs within shockwave/boundary layer interactions (SBLI) \citep{dolling2001}. There, the breathing of the TSB is associated with a low-frequency, aperiodic oscillation of a separation shock that can generate strong detrimental pressure and thermal loads on the structure, as described in the review articles by \cite{Dussauge2006}, \cite{Clemens2014}, and \cite{Gaitonde2015}. Recently, evidence of similar low-frequency unsteadiness has also been observed in subsonic flows. \cite{weiss2015} and \cite{mohammed2016} experimentally set up a TSB on a flat test surface through the combination of an adverse and a favorable pressure gradient. They observed its low-frequency breathing at a Strouhal number similar to SBLIs ($St \simeq 0.01$). Consistent findings were also reported by \cite{richardson2023} in a configuration that only featured an adverse pressure gradient (APG) but no favorable pressure gradient (FPG), by \cite{weiss2022} in a turbulent half-diffuser flow, and by \cite{dau2023} in the separation bubble behind a wall-mounted hump. On the numerical side, \cite{wu2020} computed a configuration similar to \cite{richardson2023} via direct numerical simulation (DNS) but did not capture the low frequencies observed in the experiment. On the other hand, \cite{Larcheveque2020} showed good agreement between the characteristic frequency of the breathing motion ($St \simeq 0.01$) of flat-plate TSBs at low-subsonic, high-subsonic, and supersonic flows based on large eddy simulations (LES). This suggests that similar low-frequency behaviour occurs in a wide range of Mach numbers, as argued by \cite{weiss2015,Weiss2021}.

Proposed mechanisms for the occurrence of low-frequency unsteadiness in turbulent SBLIs (and, by association, subsonic TSBs) typically consist of two main categories \citep{Clemens2014}: an upstream mechanism, whereby velocity fluctuations in the incoming boundary layer directly influence the position of the separation line and modulate the size of the TSB \citep{beresh2002} - in this case the low-frequency character of the unsteadiness is explained by the presence of very-large-scale turbulent structures that have been observed in both subsonic and supersonic boundary layers \citep{ganapathisubramani2003,ganapathisubramani2007} - and a downstream mechanism, where the low-frequency unsteadiness is caused by inherent instabilities in the TSB. In the latter hypothesis, both shear layer \citep{piponniau2009} and centrifugal instabilities \citep{priebe2016low} have been considered. \cite{mohammed2016} discussed the relevance of these hypotheses for the case of subsonic pressure-induced TSBs but could not find any conclusive evidence to select a suitable mechanism. More recently, a third, intermediate model of low-frequency unsteadiness was put forward by \cite{porter2019}, who suggested that certain large-scale, near-wall perturbations in the incoming boundary layer may drive a weakly damped global mode of the separation bubble. This model inherently implies that a combination of both upstream and downstream elements is responsible for the low-frequency unsteadiness (the presence of perturbations upstream and the global mode downstream), and \say{reconciles the debate between upstream and downstream mechanisms of separation unsteadiness} \citep{porter2019}. Such an intermediate model is also consistent with the subsonic results of \cite{mohammed2021}, who experimentally demonstrated that the low-frequency behavior in their TSB is well illustrated by a first-order low-pass filter model that converts the broadband fluctuations of the incoming turbulent boundary layer into a low-frequency, large-scale oscillation of the separation and reattachment fronts. 

Relevant modern frameworks for the study of low-frequency unsteadiness in separated flows are global linear stability analysis (GLSA) and resolvent analysis (RA). Both approaches rely on a linearization of the equation of motion around a suitable base flow. In the case of GLSA, the asymptotic behaviour of the homogeneous linear system is studied to reveal the presence of global modes of oscillations and their respective growth rate \citep{theofilis2003advances, theofilis2011global}. Positive growth rates suggest that inherent instabilities are present in the flow, which may then be qualified as an oscillator capable of sustaining self-excited oscillations without the presence of forcing. GLSA has broad applications to study the stability of laminar solutions \citep{theofilis2011global}; however, when used with linearizations around the mean solution, it may provide information on dominant oscillation frequencies \citep{barkley2006linear, schmidt2017wavepackets}, although this cannot be ensured a priori \citep{sipp2007global}. On the other hand, for globally stable flows, RA investigates the forced dynamics of the linearized flow by lumping all non-linear terms that occur in the linearization process into a forcing term \citep{mckeon2010critical, hwang2010linear, cavalieri2019wave}. The approach relies on the singular value decomposition (SVD) of the resolvent operator to identify the optimal forcing and its associated linear response. As such, RA may help identify specific zones of amplification in the flow, even in the absence of any global instability. When this occurs, such flows are typically characterized as amplifiers instead of oscillators \citep{huerre1990}.

To date, most global linear analyses of separation bubbles have been conducted at relatively low Reynolds numbers, in an attempt to describe the onset of unsteadiness and three-dimensionality in laminar separated flows. Following the seminal work of \cite{theofilis2000}, several authors have demonstrated the existence of both oscillator- and amplifier-type dynamics in low-speed, two-dimensional laminar separation bubbles (LSB) \citep[e.g.,][]{marquet2008, ehrenstein2008, rodriguez2017three}. At higher Mach numbers, the topic of low-frequency unsteadiness was specifically addressed by \cite{robinet2007}, who investigated the stability of a laminar SBLI through GLSA and demonstrated the appearance of a three-dimensional global instability when the angle of the incident shock was increased. The presence of low-frequency unsteadiness in laminar SBLIs was later confirmed by \cite{sansica2016}, who linked its appearance to the laminar/turbulent transition occurring in the separated shear layer and suggested that \say{the separation bubble acts as a low-pass spatial amplification filter}. Consistent results were recently presented by \cite{bugeat2022low}, who related the low-frequency behaviour in a laminar SBLI to the excitation of a stable global mode. The optimal gain computed by RA resembled a first-order low-pass filter, thereby recovering the signature of low-frequency unsteadiness typically observed in turbulent SBLIs \citep{poggie2015}.  

In turbulent flows, linear analyses may also be conducted by selecting the turbulent mean as base flow \citep{crow1971orderly, michalke1984survey, del2006linear, mckeon2010critical, hwang2010linear}. \cite{touber2009} performed such a study on the mean flow obtained from the LES of a turbulent SBLI and found an unstable global mode that they suggested could be linked to the observed low-frequency unsteadiness. Consistent results were later obtained by \cite{adler2018}, who also discovered a global unstable mode related to low-frequency unsteadiness in a similar flow configuration. Through the use of linear stability analysis, \cite{sasaki2021} suggested that upstream traveling acoustic waves are responsible for the low-frequency unsteadiness in a turbulent, incident shock interaction, thereby confirming a previous hypothesis by \cite{pirozzoli2006}. While all these studies proposed that turbulent SBLIs behave as oscillators, the work of \cite{sartor2015unsteadiness} on a transonic SBLI indicated an amplifier behaviour for low-frequency perturbations. The latter conclusion was obtained by performing GLSA and RA on the average flow computed by Reynolds-Averaged Navier Stokes (RANS) simulations and comparing the results with experiments. 

In this framework, the objective of the present study is to investigate the low-frequency unsteadiness of an incompressible, pressure-gradient-induced TSB by means of global stability and resolvent analysis. Specifically, we address the upstream/downstream dichotomy by examining the characteristics of global modes and their responses to external perturbations, with the aim to better understand the mechanism causing the low-frequency breathing. From a practical perspective, our motivation is three-fold: first, subsonic TSBs may serve as a reference for more complex turbulent SBLIs, where low-frequency unsteadiness is often detrimental to flight performance; second, smooth-body flow separation remains a challenge for industrial RANS turbulent modeling, which may require specific treatment for low-frequency unsteadiness; and finally, to our knowledge, such an analysis has not yet been performed in the fluid dynamics community.

The configuration that we consider is a TSB generated on a flat test surface by a combination of adverse and favourable pressure gradients. Such a flow has already been investigated experimentally by \cite{le2020} and numerically via DNS by \cite{coleman2018}. Our chosen methodology is to perform the linear analysis on a base flow consisting of the average DNS flow field and to validate the GLSA and RA results with the unsteady experimental database. As will be seen in the next sections, this cross-validation between two separate databases strongly supports the relevance of our findings. 

The article is organised as follows. In section \ref{sec:expdata} we introduce the experimental database and demonstrate the existence of low-frequency unsteadiness from both fluctuating velocity and wall-pressure data. Then, in section \ref{sec:numdata}, we describe our chosen base flow and discuss its relevance to the present investigation. The methodology for GLSA and RA is introduced in section \ref{sec:methods} and the corresponding results are presented in section \ref{sec:results}. These results are then compared to the experimental database and other studies in section \ref{sec:disc} before a conclusion is offered in section \ref{sec:concl}. Specific details about the linear analysis are provided in the appendices.

\section{Experimental database}
\begin{figure}
\centering
\includegraphics[trim={0 0 0 0},clip,width=0.8\textwidth]{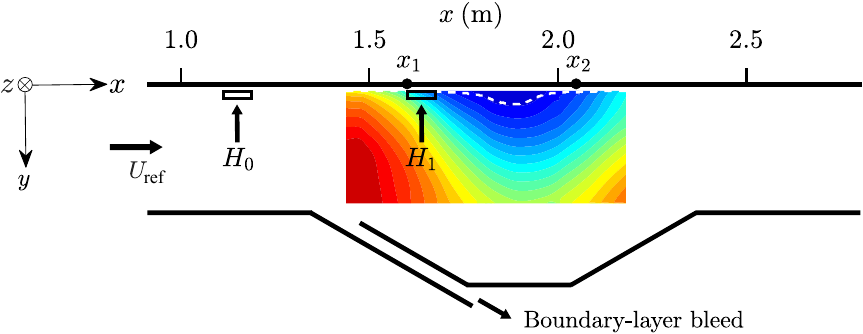}
\caption{Schematic of wind tunnel test section with  mean streamwise velocity field on the centerline. The time-averaged TSB is indicated by the dividing streamline $\overline \Psi=0$ (\dashedline). $H_0$ and $H_1$ indicate the approximate position of the near-wall TR-PIV planes.}
\label{pic:testsection}
\end{figure}
\label{sec:expdata}
In this section, we introduce the unsteady experimental database of \cite{le2018spanwise,le2020} that will be used in the present study. We briefly discuss the wind tunnel set-up, after which we proceed to outline the unsteady characteristics of the flat-plate TSB with a specific emphasis on low-frequency unsteadiness. 
\subsection{Experimental setup}
The experiments of \cite{le2018spanwise,le2020} were conducted in the TFT boundary-layer wind tunnel, a low-speed, blow-down facility specifically designed for the study of turbulent separation bubbles \citep{mohammed2015}. The wind tunnel features a test section measuring $\SI{3}{\meter}$ in length and $\SI{0.6}{\meter}$ in width. A combination of APG and FPG is generated through the widening and subsequent converging test-section floor. Whereas the APG causes the incoming zero-pressure-gradient (ZPG) flat-plate turbulent boundary layer on the upper surface to separate, a boundary-layer bleed is located on the test-section floor to ensure that the flow on the lower surface remains attached. The FPG then forces the shear layer to reattach on the upper test surface, leading to the formation of a closed turbulent separation bubble. All experiments were performed at a reference velocity $U_{\rm ref}=25$ m/s and the Reynolds number of the incoming ZPG boundary layer (based on momentum thickness) was approximately 5000.

\cite{le2020} investigated TSBs of different sizes by varying the streamwise distance between the APG and FPG. Here, we primarily focus on their medium TSB, which, as will be shown in the next section, is the closest to the flow studied numerically by \cite{coleman2018}. The length of the medium TSB, defined as the streamwise distance between the average separation and reattachment points on the test-section centerline, is $L_b = \SI{0.11}{\meter}$. The  experimental database of \cite{le2020} includes monoscopic (two-dimensional) time-resolved Particle Image Velocimetry (TR-PIV) in the streamwise/wall-normal plane as well as unsteady wall-pressure measurements on the centerline of the test section. To illustrate the spanwise character of the low-frequency unsteadiness, these results will be complemented with unsteady wall-pressure measurements in the spanwise direction by \cite{le2018spanwise} and near-wall TR-PIV measurements in the streamwise/spanwise plane of the large TSB. The latter are unpublished data from \cite{mohammed2016}. 

A schematic representation of the wind-tunnel test section, with the mean streamwise velocity field measured on the centerline, is depicted in figure~\ref{pic:testsection}. $H_0$ and $H_1$ indicate the approximate position of the near-wall TR-PIV planes, whereas $x_1$ and $x_2$ are the streamwise positions of wall-pressure measurements (see section \ref{secUnsteady}).  While the $y$-axis was oriented towards the ground during the experiments, in the remainder of the article we will switch the $y$-direction towards the top of the page. Required descriptions of the measurement techniques will be provided in the following sections as needed. More details on the experiments may be obtained in the original publications by \cite{mohammed2016} and \cite{le2018spanwise, le2020}.

\subsection{Evidence of low-frequency unsteadiness}
\label{secUnsteady}

We start by employing spectral proper orthogonal decomposition (SPOD), first introduced by \cite{lumley1970}, to characterize the low-frequency breathing of the TSB. As opposed to the \say{classical} and \say{snapshot} POD commonly found in the recent literature, SPOD produces modes that oscillate at a single frequency \citep{towne2018}. It is, therefore, a robust and powerful tool for analyzing low-frequency unsteadiness, as demonstrated for instance by the recent results of \cite{weiss2022} and \cite{richardson2023}. In practice, we use the algorithm proposed by \cite{towne2018} that is based on Welch-type averaging for stationary random processes.

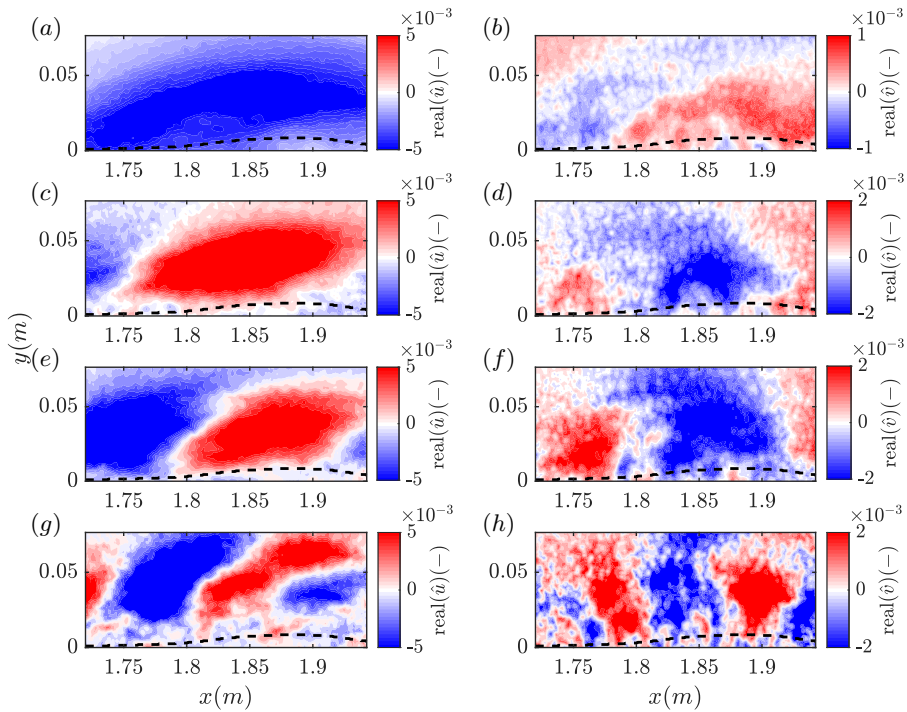
\begin{figure}
\centering

\includegraphics[trim={0 0 0 0},clip,width=\textwidth]{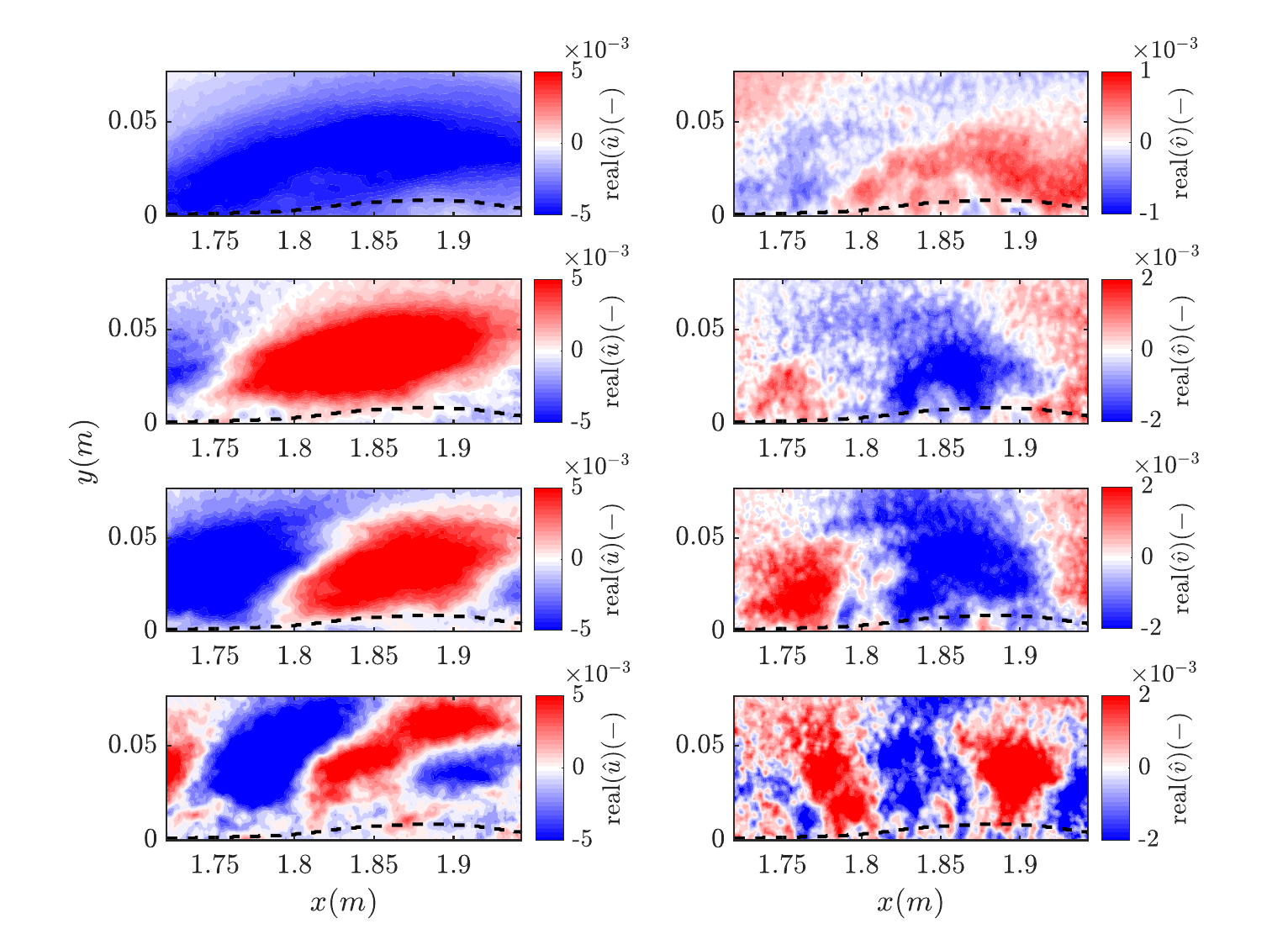}
\put(-355,262){($a$)}
\put(-185,262){($b$)}
\put(-355,200){($c$)}
\put(-185,200){($d$)}
\put(-355,138){($e$)}
\put(-185,138){($f$)}
\put(-355,76){($g$)}
\put(-185,76){($h$)}
\caption{Leading SPOD mode, computed based on monoscopic PIV measurements with $f_\textrm{s}=\SI{900}{Hz}$. The streamwise $\hat{u}$ ($a,c,e,g$) and wall-normal component $\hat{v}$ ($b,d,f,h$) are shown, respectively. The depicted frequencies, from top to bottom, correspond to $St=0.01$ $(a,b)$, $0.08$ $(c,d)$, $0.11$ $(e,f)$, $0.27$ $(g,h)$. The time-averaged location of the TSB is indicated by the dividing streamline $\overline \Psi=0$ (\dashedline).}
\label{fig:SPODmodes}
\end{figure}

In figure \ref{fig:SPODmodes} we display the streamwise and wall-normal components $\hat u$ and $\hat v$ of velocity fluctuations of the leading SPOD mode, computed from the TR-PIV measurements of \cite{le2020}. The PIV field-of-view is $\SI{225}{\milli \meter}$ x $\SI{75}{\milli \meter}$ ($x-y$) with a sampling frequency of $f_s=\SI{900}{\hertz}$. The database consists of six successive time series of $N_t=3580$ snapshots each. Hence, the decomposition is performed on a total of $N_t=21480$ PIV snapshots that are split in $72$ blocks of $N_\textrm{FFT}=512$ snapshots with $\SI{50}{\percent}$ overlap. Overlapping blocks between two consecutive (uncorrelated) runs are removed. This procedure results in a frequency resolution of $\SI{1.76}{\hertz}$.

The real part of the leading mode is depicted for different Strouhal numbers $St=fL_b/U_{\rm ref}$. A Strouhal number $St=0.01$, typically representing the low-frequency regime, is depicted on the top ($a,b$), whereas the remaining Strouhal numbers increase from top to bottom according to $St=0.08$ $(c,d)$, $St=0.11$ $(e,f)$, and $St=0.27$ $(g,h)$. As will be discussed later (see figure \ref{fig:SPOD_Eigs}), the first SPOD mode is particularly dominant at low frequency, with approximately $72\%$ of the energy density at $St = 0.01$. Note that the time-averaged location of the TSB is indicated by the dividing streamline (black-dashed line) in the different plots. 

In the low-frequency regime ($St= 0.01$) the streamwise component of the mode $\hat u$ features a large coherent structure that bounds the TSB and follows its shape. This behavior can be observed for any low Strouhal number with $St\approx 0.01$. A similar, large-scale mode was first observed in the snapshot POD of the streamwise velocity component in \cite{mohammed2016}. In their study, a low-order model was employed to show that this mode can be interpreted as a low-frequency contraction and expansion (breathing) of the TSB. More recently, SPOD has been applied to several pressure-induced TSB flows and large-scale coherent structures similar to that depicted in figure \ref{fig:SPODmodes} (a,b) have been identified as the leading low-frequency SPOD mode for TSBs occurring in a one-sided diffuser \citep{steinfurth2022}, behind a wall-mounted hump \citep{dau2023} and on a flat-plate with APG \citep{richardson2023}. In all of these works, the aforementioned mode was associated with the low-frequency breathing of the TSB. In contrast, higher-frequency modes are qualitatively distinct: There, we observe an alternating pattern of coherent structures of opposite phase, which, in the literature, is often associated with the shedding of vortices from the shear layer bounding the recirculation region \citep[e.g.,][]{rajaee1994}. The corresponding wall-normal component of the modes conveys similar trends. While in the low-frequency regime, we observe structures that encompass a significant portion of the PIV domain, an increase in the Strouhal number results in both a greater number of structures as well as smaller individual structure size.

Figure \ref{fig:SPODmodes} (a) indicates that, at low frequency, the TSB contracts and expands in the streamwise direction \citep[see also the discussion in the original article by][]{le2020}. In order to estimate the spanwise scale of this motion, we now consider TR-PIV measurements performed in the streamwise/spanwise plane at an elevation of $y \simeq 4$~mm away from the wall (for reference, the boundary layer thickness at position $H_0$ is approximately 30~mm.) These measurements were performed in the large TSB of \cite{mohammed2016} (unpublished data, see also \cite{MTaifour2017}). The PIV field-of-view is $\SI{75}{\milli \meter}$ x $\SI{215}{\milli \meter}$ ($x-z$) with a sampling frequency of $f_s=\SI{900}{\hertz}$. 

\begin{figure}
\centering
\includegraphics[trim={0cm 0cm 0cm 0cm},clip,width=0.9\textwidth]{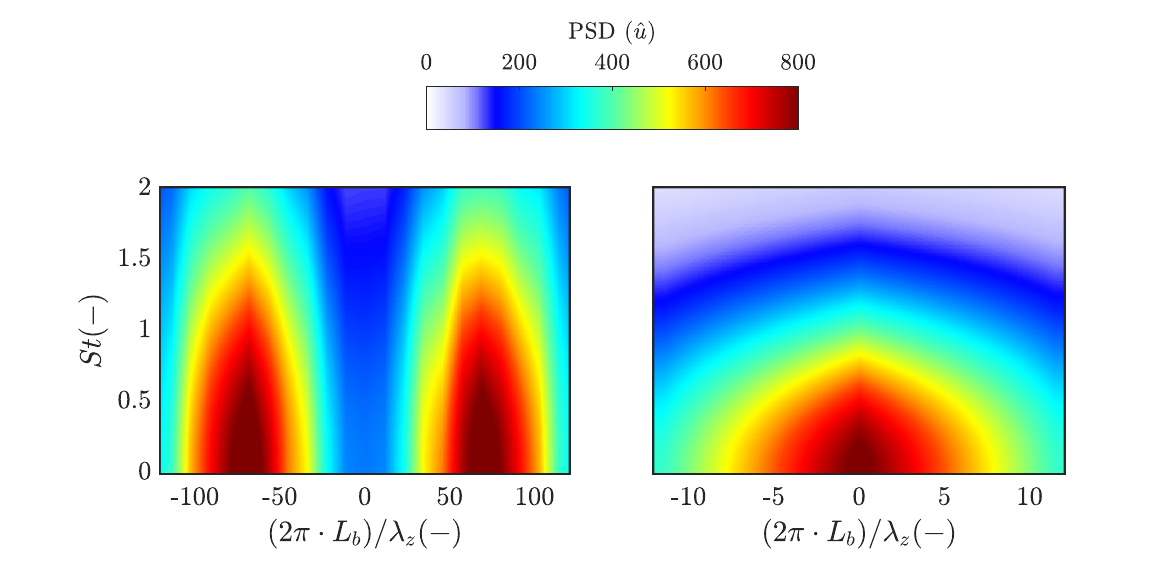}
            \put(-320,120){($a$)}
            \put(-170,120){($b$)}
\caption{Frequency-wavenumber spectrum of fluctuating velocity $u'$ in the $x-z$ plane at $y\simeq 4$~mm from the test surface. Two streamwise positions $x$ are depicted, a position in the ZPG region upstream of the TSB $x=\SI{1.33}{\meter}$ ($a$) and a position immediately upstream of the time-averaged location of the separation line $x=\SI{1.68}{\meter}$ ($b$). The spanwise wavenumber $\beta=2\pi/\lambda_\textrm{z}$ is non-dimensionalized by the average bubble length $L_b$. The PSD in the ZPG region ($a$) is multiplied by a factor of $10$.} 
\label{fig:FFTu}
\end{figure}

The frequency-wavenumber spectra of the streamwise velocity fluctuations is depicted in  figure \ref{fig:FFTu}. Two exemplary streamwise position are chosen, representing the spanwise ($x-z$) planes $\textrm{$H_0$}$ (Fig. \ref{fig:FFTu} $(a)$) and $\textrm{$H_1$}$ (Fig. \ref{fig:FFTu} $(b)$). Here, $\textrm{$H_0$}$ is a plane in the ZPG region upstream of the TSB and $\textrm{$H_1$}$ corresponds to a plane immediately upstream of the time-averaged location of the separation line, see figure \ref{pic:testsection}. To extract the frequency-wavenumber spectra, we first perform a Fast Fourier Transform (FFT) in the spanwise ($z$) direction on $N_t=3580$ snapshots of the fluctuating velocity $u'$ \citep[e.g.,][]{towne2017}. We further compute the power spectral density (PSD) of the Fourier-transformed signal $\hat u$ adopting Welch’s estimate using $\SI{50}{\percent}$ overlap. We obtain a spectrum for each streamwise position $x$. Here, the frequency is represented as the Strouhal number $St=f L_b/U_{\rm ref}$ and the non-dimensional spanwise wavenumber is $\beta=(2\pi \cdot L_b)/\lambda_z$. The chosen position corresponds to the center of each PIV measurement plane, respectively. Similar results were obtained for all investigated positions within the planes $\textrm{$H_0$}$ and $\textrm{$H_1$}$. For visualization purposes, the PSD of $\hat u$ is scaled by a factor of $10$ for the $\textrm{$H_0$}$ plane in the upstream boundary layer (Fig. \ref{fig:FFTu} $(a)$). 

In the upstream ZPG boundary layer ($H_0$ plane), the maximum of the PSD of $\hat u$ is obtained for the non-dimensional spanwise wavenumber $\beta=\pm 65$ and a Strouhal number range of $St\le0.8$ (Fig. \ref{fig:FFTu} ($a$)). All frequencies $St$ in the vicinity of $\beta=\pm 65$ exhibit fairly high energy levels. However, no substantial energy content can be detected in the region of two-dimensional perturbations $\beta=0$. Moreover, since the spectrum is symmetrical with respect to $\beta=0$, no preferential $z$-direction can be detected for the present configuration. This indicates that perturbations are equally likely to propagate in the $+z$ and $-z$ direction. When expressed in terms of the boundary layer thickness $\delta=28$~mm of the incoming turbulent flow (measured at $x=\SI{1.1}{\meter}$), the spanwise wavelength $\lambda_z$, related to the non-dimensional spanwise wavenumber $\beta=\pm 65$, is approximately $\lambda_z=0.4\delta$. This value is in agreement with the values of $\mathcal{O}(\delta)$ reported for the superstructures in the upstream boundary layer of several turbulent boundary layer flows \citep[e.g.,][]{tomkins2003spanwise, hutchins2007evidence, le2016effect}. Interestingly, a similar organization of elongated structures over the span of the  $\textrm{$H_0$}$ measurement plane can also be observed on the first SPOD modes at low frequency (not shown here).

In figure \ref{fig:FFTu} ($b$) we display the representative frequency-wavenumber spectrum of the measurement-plane $\textrm{$H_1$}$. The majority of the energetic content is now gathered in the range $-2\le\beta\le2$. Once again, the distribution is symmetrical with respect to $\beta=0$, indicating no preferential $z$-direction. On the other hand, the wavelengths $\lambda_z$ associated with the structures of this low-$\beta$ range are now large compared to the bubble length $L_b$. They take values between $\lambda_z=3L_b$ for $\beta=\pm2$ to $\lambda_z\gg 10L_b$ for very small $\beta$. Hence, the low-frequency unsteadiness of the TSB appears to be coherent over a large spanwise scale. Furthermore, the high-$\beta$ signature observed in the incoming ZPG boundary layer is absent from the frequency-wavenumber spectrum close to the separation line, thereby suggesting that the TSB is not directly responding to the long superstructures present in the incoming boundary layer.

To confirm these results, we now consider fluctuating wall-pressure data gathered in the spanwise direction by \cite{le2018spanwise} in the medium TSB again. All pressure signals were obtained by using piezoresistive pressure transducers with a range of $1$ psi ($\SI{6.89}{\kilo \pascal})$ and an estimated error of $\pm \SI{5}{\percent}$. To eliminate low-frequency wind-tunnel noise in the signals, the correction method from \cite{weiss2015} was applied. PSDs were calculated by adopting Welch’s method using $\SI{50}{\percent}$ overlap and a Hamming window. 

\begin{figure}
\centering
\includegraphics[trim={0 0 0 0},clip,width=\textwidth]{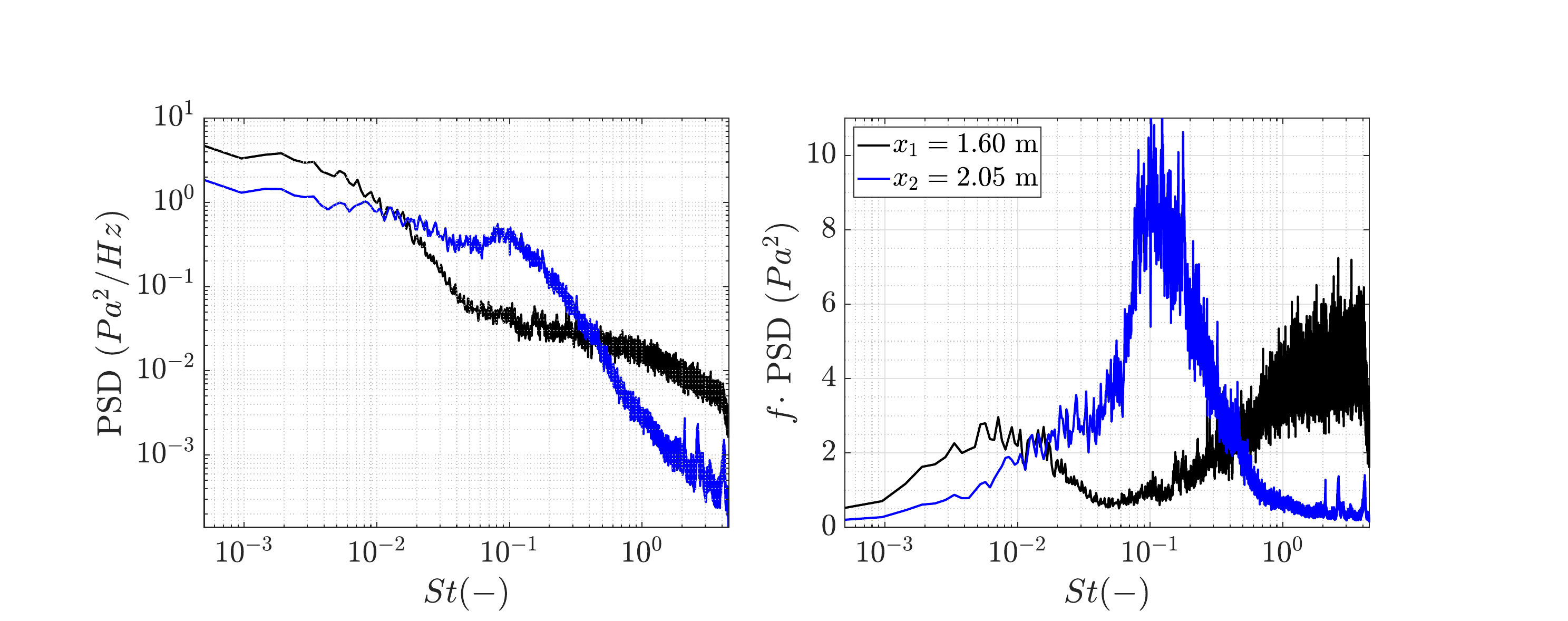}
            \put(-360,132){($a$)}
            \put(-195,132){($b$)}
\caption{Power spectral density (PSD) ($a$) and pre-multiplied PSD ($b$) of fluctuation pressure $p'$ for $x_1=\SI{1.60}{\meter}$ (black) and $x_2=\SI{2.05}{\meter}$ (blue). The Strouhal number $St$ is calculated based on the bubble length $L_b$ and the reference velocity $U_\textrm{ref}=\SI{25}{\meter \per \second}$.}
\label{pic:Pressure}
\end{figure}

In figure \ref{pic:Pressure}, the classical log-log PSD ($a$) and the pre-multiplied PSD ($b$) of the fluctuating wall pressure $p'$ on the test-section centerline at two streamwise positions are shown: a position immediately upstream of the time-averaged location of the separation bubble ($x_1=\SI{1.60}{\meter}$) and a position in the region downstream of the TSB ($x_2=\SI{2.05}{\meter}$). These streamwise positions are depicted schematically in figure \ref{pic:testsection}. The low-frequency unsteadiness becomes apparent as a distinct \say{hump} in the pre-multiplied distribution for $x_1=\SI{1.60}{\meter}$, where a significant amount of energy is gathered in the region $St\approx 0.01$. This hump has also been observed by \cite{mohammed2016} and \cite{richardson2023}, and has been associated to the low-frequency breathing of their TSB. On the other hand, a different behavior can be observed for  $x_2=\SI{2.05}{\meter}$. Here, a distinct peak is visible in the pre-multiplied PSD for $St\approx 0.1$. In \cite{cura2023medium}, this medium-frequency unsteadiness was linked to instabilities in the shear layer bounding the recirculation region.

\begin{figure}
\centering
\includegraphics[trim={0 0 0 0},clip,width=\textwidth]{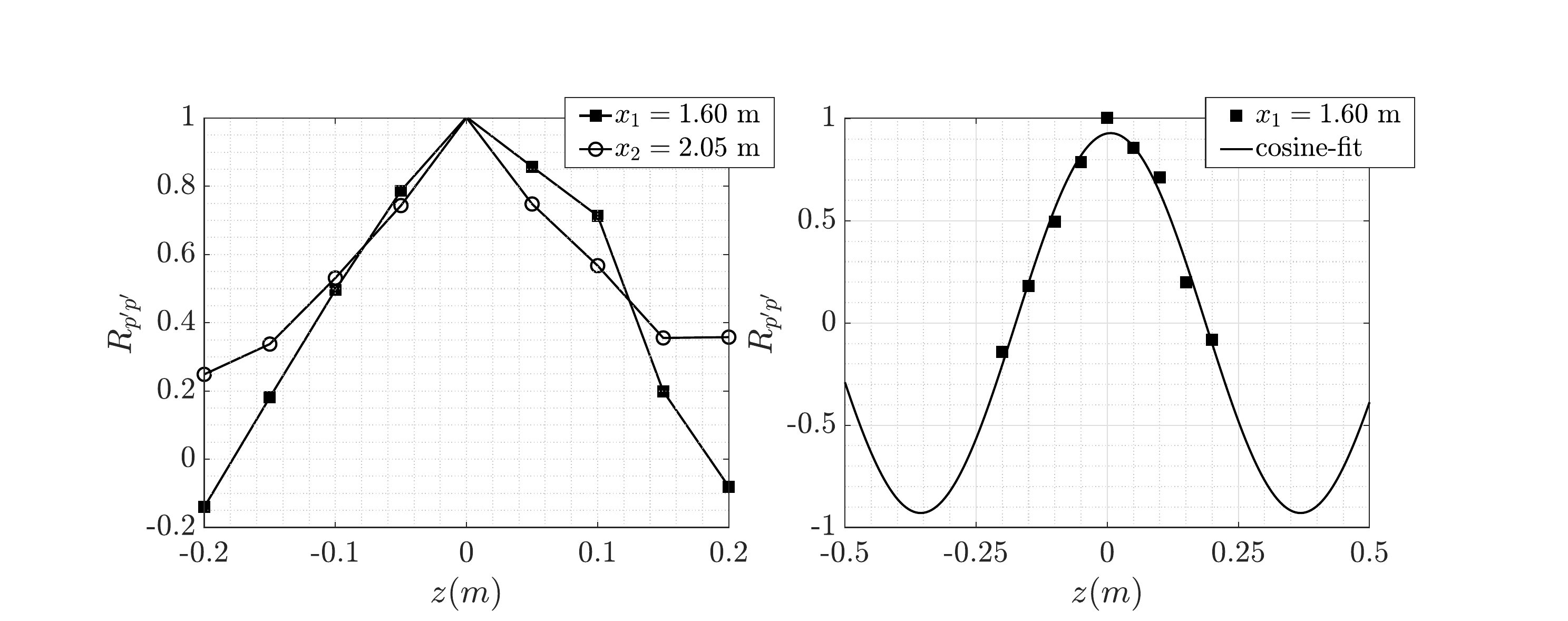}
            \put(-352,132){($a$)}
            \put(-195,132){($b$)}
\caption{Spanwise correlation $R_{p^\prime p^\prime}$ of low-pass filtered fluctuating pressure for two streamwise positions  $x_1=\SI{1.60}{\meter}$ (\solidlinesquare)
, $x_2=\SI{2.05}{\meter}$ (\solidlinecirc) modified from \cite{le2018spanwise} ($a$)  and cosine-fit $f(z)=c_1\cdot \textrm{cos}(c_2z)$ of  $R_{p^\prime p^\prime}$ (\solidline) at  $x_1=\SI{1.60}{\meter}$ ($b$). The resulting non-dimensional spanwise wavenumber $\beta= (2\pi \cdot L_{b})/\lambda_\textrm{z}$ is equal to $0.97$. The fluctuating pressure at the test-section centerline ($z=\SI{0}{m}$) is used as reference for all correlations.}
\label{pic:Rpp}
\end{figure}

Coming back to the spanwise characteristics of low-frequency unsteadiness, we show in figure \ref{pic:Rpp} ($a$) the two-point cross-correlation coefficient at zero time lag measured by \cite{le2018spanwise}. Here, $R_{p'p'}=\overline{p'(z)*p'_\textrm{ref}(z_\textrm{ref})}/(p'_\textrm{rms}*p'_\textrm{ref,rms})$ was obtained by simultaneously measuring the wall-pressure fluctuations at the centerline of the test section ($z_\textrm{ref}=0$~mm) and with a moving sensor positioned successively at $z=[0, \pm0.05, \pm0.10, \pm0.15, \pm0.20]$~m along the span of the wind-tunnel test section. Furthermore, all pressure signals were low-pass filtered to frequencies below $St = 0.03$ before computing the cross-correlations, so that only values that correspond to the low-frequency hump in figure \ref{pic:Pressure} are considered. Again, two streamwise positions,  $x_1=\SI{1.60}{\meter}$ and  $x_2=\SI{2.05}{\meter}$, are depicted. In both cases, a wave-like distribution of $R_{p'p'}$ over the span, with a relatively large wavelength $\lambda_z$, becomes apparent. This confirms that the low-frequency unsteadiness is coherent over a large portion of the test-section span.

To obtain a quantitative metric of spanwise coherence, we now perform a curve-fit of the correlation curve at  $x_1=\SI{1.60}{\meter}$,  using a cosine function of the form
\begin{equation}
    f(z)=c_1\cdot \textrm{cos}(c_2z).
\end{equation}
We obtain the distribution shown in figure \ref{pic:Rpp} ($b$), where the cosine-function has a non-dimensional spanwise wavenumber $\beta=0.97$. This value is in the energy-containing range observed in the frequency-wavenumber spectrum of near-wall velocity data in figure \ref{fig:FFTu} ($b$). Notably, this results closely matches the spanwise wavenumber corresponding to the width of the wind tunnel $b=\SI{0.6}{\meter}$, which is $\beta= 1.17$.

In summary, the experimental results obtained by \cite{le2018spanwise, le2020}, using both fluctuating velocity and wall-pressure measurements, indicate that the TSB is contracting and expanding at low frequency, with a characteristic Strouhal number of the order of $St=fL_b/U_{\rm ref} = 0.01$. This breathing motion appears to be reasonably coherent across the span, with a spanwise wavenumber of the order of $\beta=(2\pi \cdot L_b)/\lambda_z = 1$. In the remainder of the article, our main objective will be to use global linear stability and resolvent analysis to try and explain the origin of this motion.
\section{Numerical database}
\label{sec:numdata}
In this work, we perform modal and non-modal stability analysis of the time- and spanwise averaged velocity field from the DNS by \cite{coleman2018}. Our motivation in doing so is mainly that employing an experimental flow field as base flow for stability analysis would typically require some degree of curve-fitting and/or extrapolation (e.g. \cite{nishioka1990control, yarusevych2006coherent}). This is mostly due to the inherent characteristics of experimental velocity fields, which typically exhibit considerable data scatter. Both the application of curve-fitting techniques and the presence of data scatter are well-documented phenomena known to significantly influence the outcomes of local linear stability analysis \citep[e.g.,][]{dovgal1994, bottaro2003effect, boutilier2013}. Furthermore, velocity fields measured by TR-PIV typically suffer from poor spatial resolution because of the relatively large pixel size of CMOS cameras \citep{le2020}. On the contrary, a DNS base flow usually provides a higher spatial resolution and a larger domain size than its experimental counterpart, hence facilitating the computation of the required derivatives in the stability analysis. 

A typical GLSA/RA study on a DNS base flow would usually be compared to unsteady DNS data. Here, we do not follow this path but we compare our results to the experimental database of \cite{le2018spanwise,le2020} summarized in the last section. Our motivation for doing so is two-fold. First, \cite{coleman2018} could not observe any low-frequency breathing motion in their DNS data because of limited computing resources. Indeed, capturing low-frequency unsteadiness requires very long integration times that are still difficult to reach for well-resolved simulations. Similarly, \cite{wu2020} were not able to capture $St\simeq 0.01$-phenomena in their own DNS. Hence, to the authors' knowledge, the low-frequency unsteadiness has not been observed in any simulation of low-speed TSB yet. Our second motivation is that, if the linear analysis on the DNS base flow favorably compares to the experimental results, it would strongly support the generality and portability of our results, and also rule out that the breathing is caused by an experimental artifact. Therefore, although the DNS of \cite{coleman2018} was not originally designed to match our experimental TSB flow, we contend that the merits of our strategy outweigh its drawbacks. 

The flow field of \cite{coleman2018} was selected because of its similarity to the experimental set-up and its free accessibility \citep{DNS}. The DNS features a fully turbulent, two-dimensional flat-plate boundary layer, which is subjected to an APG subsequently followed by an FPG. The APG-FPG characteristics are enforced by a transpiration velocity profile $V_\textrm{top}(x)$, introduced on a virtual parallel plane at a fixed distance opposite to the no-slip wall (Eq. \ref{eq1}). Here, $V_\textrm{max}$ is the maximum velocity of the transpiration velocity profile and $\xi$ is the length scale. To enforce the ZPG along the wall, a constant boundary-layer bleed velocity $\phi_\textrm{top}$ is introduced. A pseudo-spectral code was used to compute the solution to the incompressible Navier-Stokes equations. Further details regarding the DNS can be found in \cite{coleman2018}. In the following, only the mean flow field from case C (main case) will be considered.
\begin{equation}
V_\textrm{top}(x)=-\sqrt{2}V_\textrm{max}\Bigl(\frac{x}{\xi}\Bigr)\textrm{exp}\Bigl(\frac{1}{2}-\Bigl(\frac{x}{\xi}\Bigr)^2\Bigr)+\phi_\textrm{top}
\label{eq1}
\end{equation}
The similarities between the DNS (Case C) and the experimental (medium) TSB were already outlined in \cite{le2020}. Here, we demonstrate that the two TSB flows exhibit an even higher degree of similarity when an appropriate set of scaling parameters is selected. For this purpose, we introduce the parameter $L_p$, which is equal to the distance between the maximum APG and FPG. We further introduce the momentum thickness $\Theta_0$, which was computed by means of the von Kármán integral $d\Theta/dx=c_f/2$. It correspond to the momentum thickness that would be reached at the streamwise position $x(V_\textrm{top}=0)$ for a ZPG boundary layer \citep{coleman2018}.

\begin{figure}
\centering
\includegraphics[trim={0 0 0 0},clip,width=.8\textwidth]{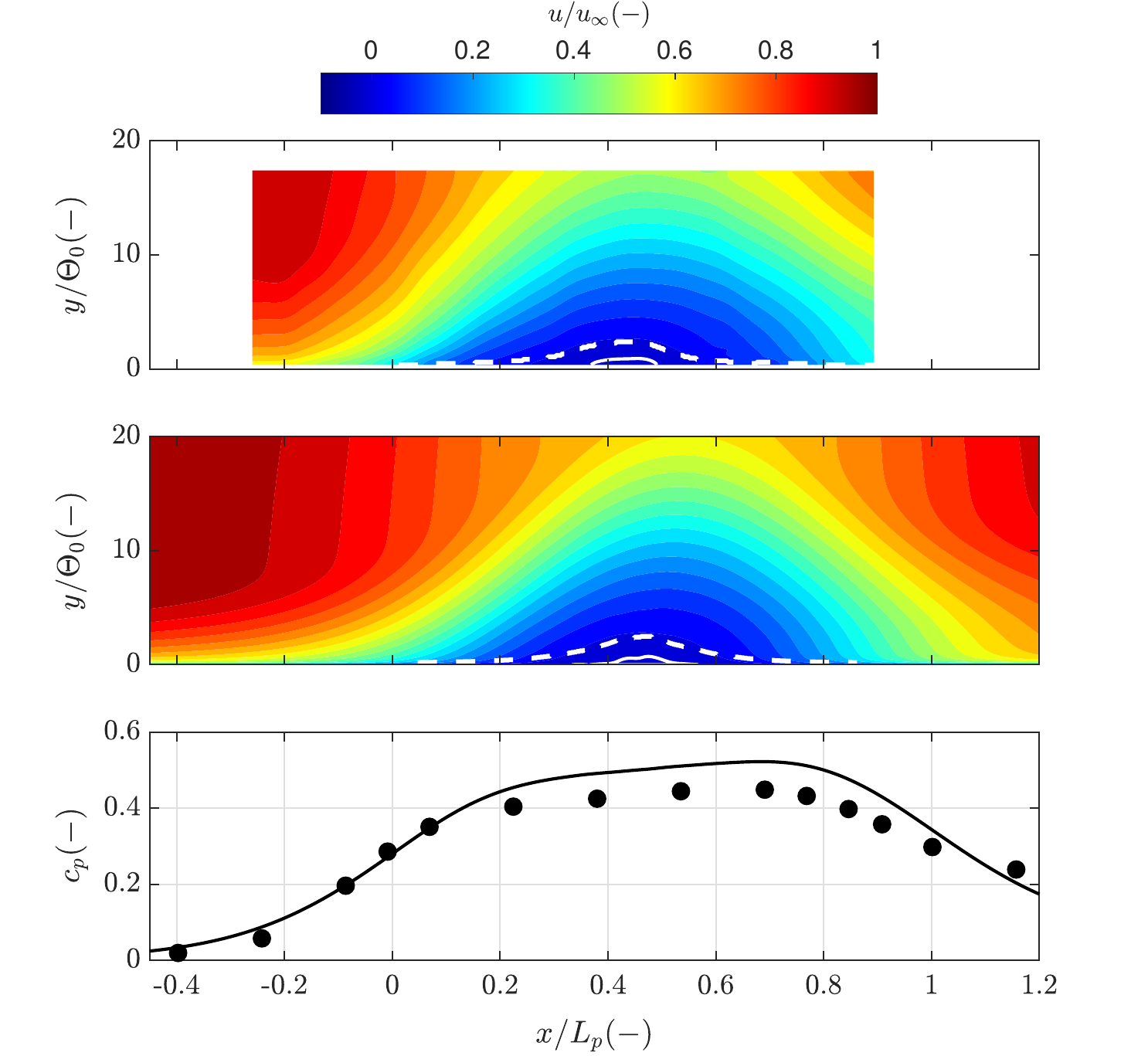}
            \put(-290,252){($a$)}
            \put(-290,173){($b$)}
             \put(-290,95){($c$)}
\caption{Streamwise velocity component $u/u_\infty$ of experimental flow field ($a$) and DNS base flow ($b$). The pressure distribution along the flat plate is represented by the pressure coefficient $c_p$ ($c$) for the DNS base flow (\solidline) and the experimental data (\circlesymb ). The streamwise and wall-normal coordinates are non-dimensionalized by $L_p$ and $\Theta_0$, respectively. We indicate the time-averaged position of the TSB by the $\overline{u}=0$ contourline (\solidline) and the dividing streamline (\dashedline). }
\label{picFlows}
\end{figure}

In figure \ref{picFlows} we plot the streamwise velocity $u/u_\infty$ of the experimental flow from \cite{le2020} ($a$) and the DNS calculations from \cite{coleman2018} ($b$). The streamwise and wall-normal coordinates are non-dimensionalized by the distance $L_p$ and the momentum thickness $\Theta_0$, respectively. For each flow, we indicate the time-averaged location of the TSB by means of the $\overline{u}=0$ contourline (white-solid line) and the dividing streamline (white-dashed line). In this scaling, the flow databases become very similar, with a comparable position of mean flow separation and reattachment. 
\begin{table}
\centering
\begin{tabular}{lcccc}
Database & $Re_{\Theta_0} (-)$ & $Re_{L_b} (-)$ & $L_b/\Theta_0(-)$ & $L_p/\Theta_0(-)$ \\ 
\hline
Medium TSB (\cite{le2020}) &  $6905$ &  $183673$ & $26.6$ & $154.5$\\ 
Case C (\cite{coleman2018}) &  $3121$ & $104000$ & $33.3$ & $207.5$\\ 
\end{tabular}
\label{tab1}
\caption{Flow characteristics of experimental and numerical database.}
\end{table} 
 In figure \ref{picFlows} ($c$), the pressure distribution along the flat plate, as indicated by the pressure coefficient $c_p$, is displayed. The solid line represents the DNS, whereas symbols pertain to the wall-pressure measurements. In contrast to the contour plots of streamwise velocity $u/u_\infty$, some discrepancies between the DNS base flow and the experimental data become evident. The plateau in the distribution of $c_p$ is reached for a higher value in the case of the DNS. However, up to $x/L_p\approx 0.9$, the distributions remain in good qualitative agreement. 
 While acknowledging the non-identical nature of both flows, we operate under the assumption that insights derived from the linear analysis of the DNS base flow can be transferred to the experimental counterpart. This assumption will be re-evaluated in the discussion section of this paper. The key characteristics of each flow database are summarized in table \ref{tab1}.
\section{Methodology}\label{sec:methods}
In this section, the governing equations describing the dynamics of the turbulent separation bubble (TSB), as well as the employed stability and data analysis methods, are described.
\subsection{Governing equations}
\begin{figure}
\centering
\includegraphics[trim={0 0 0 0},clip,width=.8\textwidth]{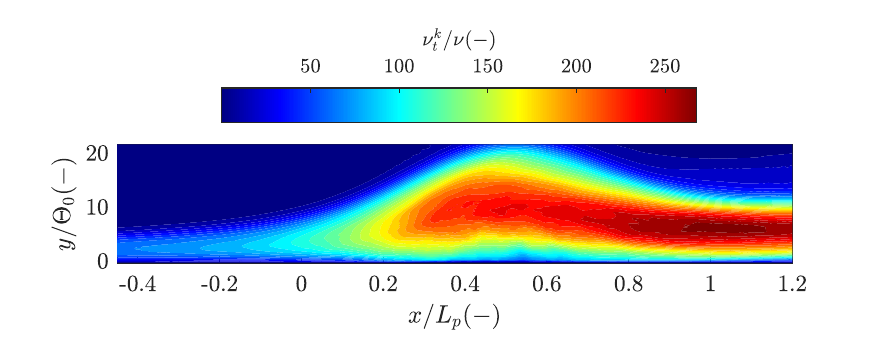}
\caption{Eddy viscosity calculated from DNS data.}
\label{picnut}
\end{figure}

The viscous incompressible Navier-Stokes equations for the conservative variables $\textbf{q}=(u,v,w,p)$
\begin{equation}
\nabla \cdot \textbf{u}=0, \qquad \frac{\partial \textbf{u}}{\partial t}+\textbf{u}\cdot \nabla \textbf{u} =-\nabla p + \frac{1}{Re}\nabla ^2 \textbf{u}, 
\label{eqNST}
\end{equation}
are considered, where $\textbf{u}=(u,v,w)$ are the streamwise, wall-normal and spanwise velocity, respectively, $p$ is the pressure, and $Re$ is the Reynolds number. Here, we decompose the flow field into time-averaged and fluctuating quantities according to
\begin{equation}
 \textbf{q}(x,y,z,t)=\overline{\textbf{q}}(x,y,z)+\tilde{\textbf{q}}(x,y,z,t). 
 \label{eqReynoldsDecomp}
\end{equation}
Introducing the above Reynolds decomposition (Eq. \ref{eqReynoldsDecomp}) into the incompressible Navier-Stokes equations (\ref{eqNST}) and time-averaging yields the linearized Navier-Stokes equations (LNSE)
\begin{equation}
\centering
\frac{\partial \tilde{\textbf{u}}}{\partial t}+\tilde{\textbf{u}}\cdot\nabla\overline{\textbf{u}}+\overline{\textbf{u}}\cdot\nabla\tilde{\textbf{u}}=-\nabla\tilde{p}+\frac{1}{\textrm{Re}}\nabla^2\tilde{\textbf{u}}+\tilde{\textbf{f}},
\label{LNSE_turb}
\end{equation}
\begin{equation}
\centering
\nabla\cdot \tilde{\textbf{u}}=0.
\label{LNSE_cont}
\end{equation}
Here, we group the non-linear terms in the Navier-Stokes equations into a forcing term $\tilde{\textbf{f}}$, as proposed by \cite{mckeon2010critical}. In the work of \cite{towne2018} it was demonstrated that when the forcing term is modelled as spatial white noise, a direct relationship between SPOD and resolvent modes can be expected. However, turbulent flows have non-linear terms that differ from such white-noise approximation, and thus have colour \citep{zare2017colour, morra2021colour, nogueira2021forcing}. When linearizing around a turbulent mean flow, the colour of the forcing is often partly incorporated by means of an eddy-viscosity model \citep{morra2019relevance, morra2021colour}. Here, we follow the methodology outlined in \cite{reynolds1972} and represent part of the Reynolds stresses by means of an eddy-viscosity model
\begin{equation}
\frac{\partial \tilde{\textbf{u}}}{\partial t}+\tilde{\textbf{u}}\cdot\nabla\overline{\textbf{u}}+\overline{\textbf{u}}\cdot\nabla\tilde{\textbf{u}}=-\nabla\tilde{p}+\frac{1}{\textrm{Re}}(1+\nu_t/\nu)\nabla^2\tilde{\textbf{u}}+\tilde{\textbf{f}}.
\label{eqrij2}
\end{equation}
The eddy viscosity is calculated from the DNS data provided in \cite{coleman2018} as $\nu_{t}=c_\mu{k^2}/{\varepsilon}$ (see Fig. \ref{picnut}). Here $c_\mu=0.09$, $k$ is the turbulent kinetic energy, and $\varepsilon$ is the dissipation rate.
%

\subsection{Global linear stability analysis}
When the forcing term is set to $\tilde{\textbf{f}}=0$, the system of equations (\ref{LNSE_turb}-\ref{LNSE_cont}) can be recast in matrix form as
\begin{equation}
\textbf{M} \frac{\partial \tilde{\textbf{q}}}{\partial t}=\textbf{A}_{3D}\tilde{\textbf{q}},
\end{equation}
where $\textbf{A}_{3D}$ is the three-dimensional LNSE operator.
Choosing a modal ansatz of the form
\begin{equation}
\tilde{\textbf{q}}(x,y,z,t)=\hat{\textbf{q}}(x,y,z)\textrm{exp}( -i\omega t)+c.c.,
\label{eqAnsatz}
\end{equation}
and introducing it into the LNSE (\ref{LNSE_turb}-\ref{LNSE_cont}) leads to a generalized eigenvalue problem (EVP)
\begin{equation}
    -i\omega \textbf{M}\hat{\textbf{q}}(x,y,z)=\textbf{A}_{3D}(\overline{\textbf{q}}(x,y,z),Re)\hat{\textbf{q}}(x,y,z),
    \label{eqEVP}
\end{equation}
where c.c. is the complex conjugate, $\omega \in \mathbf{C}$ are the eigenvalues, and $\hat{\textbf{q}}$ are the eigenfunctions. We now assume homogeneity in the spanwise direction $z$ and perform a Fourier transform in space, such that equation (\ref{eqAnsatz}) reduces to
\begin{equation}
\tilde{\textbf{q}}(x,y,z,t)=\hat{\textbf{q}}(x,y)\textrm{exp}[i( \beta z-\omega t)]+c.c.,
\label{eqAnsatz2}
\end{equation}
and the EVP can be reformulated as
\begin{equation}
    -i\omega \textbf{M}\hat{\textbf{q}}(x,y)=\textbf{A}_{2D,z}(\overline{\textbf{q}}(x,y),\beta,Re)\hat{\textbf{q}}(x,y).
    \label{eqEVP2D}
\end{equation}
Here, $\beta \in \mathbf{R}$ is the spanwise wavenumber. The two-dimensional LNSE operator $\textbf{A}_{2D,z}$ can be extracted from $\textbf{A}_{3D}$ by introducing $\overline{\textbf{u}}=(\overline{u},\overline{v},0)$ and employing equation (\ref{eqAnsatz2}).  The operator can then be divided into $\textbf{A}_{2D,z}=\textbf{A}+\textbf{N}$, where $\textbf{N}$ represents the (additional) turbulence terms modeled in equation (\ref{eqrij2}). The EVP in equation (\ref{eqEVP2D}) needs to be supplemented with appropriate homogeneous boundary conditions to fulfill the physical constraints on the domain (see Appendix \ref{appB}). The spanwise-periodic EVP can then be analyzed to yield unstable global modes whenever the growth rate $\omega_i>0$, whereas disturbances decay in the asymptotic time limit for $\omega_i<0$.

The LNSE operator $\textbf{A}_{2D,z}$ and operator $\textbf{M}$ are included in Appendix \ref{appA}. The solution of the 2D EVP is performed using the code presented in \cite{abreu2021spanwise} and \cite{blanco2022improved}, adapted for global stability analysis. A grid and fringe convergence study was performed and is summarized in Appendix \ref{appC}. All quantities appearing in the linear analysis are non-dimensionalized by means of the length scale $l^*=L_b$ and the time-scale $t^*=L_b/u_\infty$. The resulting Reynolds number based on $u_\infty$ and $L_b$ is $Re=104 000$. On the other hand, in order to plot the results in a manner consistent with the experiments, the streamwise and wall-normal coordinates of the TSB are represented as $x/L_p$ and $y/\Theta_0$, respectively.
\subsection{Resolvent analysis}
In order to study the linear forced dynamics of the system, we now consider equation (\ref{eqEVP2D}) and re-introduce the forcing term on the right-hand-side
\begin{equation}
-i\omega\textbf{M} \hat{\textbf{q}}=\textbf{A}_{2D,z}\hat{\textbf{q}}+\textbf{B}\hat{\textbf{f}},    
\end{equation}
where the harmonic forcing is $\hat{\textbf{f}}=(f_x,f_y,f_z,0)$. This system can be rewritten in the resolvent form
\begin{equation}
    (-i\omega \textbf{M}-\textbf{A}_{2D,z})\hat{\textbf{q}}=\textbf{B}\hat{\textbf{f}}.
\end{equation}
The optimal response $\hat{\textbf{q}}$ to any harmonic forcing $\hat{\textbf{f}}$ can be obtained by performing a singular value decomposition (SVD) of the resolvent operator $\textbf{R}$
\begin{equation}
\hat{\textbf{q}}=\textbf{C}(-i \omega \textbf{M}-\textbf{A}_{2D,z})^{-1}\textbf{B}\hat{\textbf{f}}=\textbf{R}\hat{\textbf{f}},
\label{eq:resolvent}
\end{equation}
where the operators $\textbf{B}$ and $\textbf{C}$ act as filters that impose restrictions on the forcing (input) and the response (output), respectively. 
The first singular value of the SVD of the resolvent operator is then the optimal gain $\sigma_1$, whereas the left and right singular vectors represent the optimal forcing and response, respectively. The remaining singular values of the SVD are called the sub-optimal gains, and are arranged in decreasing order $\sigma_1>\sigma_2>\sigma_3>...>\sigma_n$. The solution to the SVD in equation (\ref{eq:resolvent}) is computed using the code presented in \cite{abreu2021spanwise} and \cite{blanco2022improved}. The operators $\textbf{B}$ and $\textbf{C}$ are discussed in more detail in Appendix \ref{appA}.
The study conducted by \cite{towne2018} demonstrated a direct relationship between the optimal response obtained from RA and the modes extracted from SPOD, under the assumption that the forcing is modeled as spatial white noise. Even though non-linearities in the Navier-Stokes equations are expected to have \say{colour} \citep{zare2017colour}, a strong link between RA and SPOD can be expected if the CSD is dominated by the optimal response  \citep{cavalieri2019wave}. This is the case when the leading resolvent mode is of low-rank, such that $\sigma_1\gg\sigma_2$.

In order to quantify the alignment between SPOD and RA \citep{lesshafft2019resolvent, abreu2020spectral} for several spanwise wavenumbers $\beta$, we introduce the metric
\begin{equation}
    \varphi=\frac{\Bigl\langle \hat q_{1_\textrm{SPOD}}, \hat q_{1_\textrm{RA}}\Bigr\rangle}{\|  \hat q_{1_\textrm{SPOD}}\| \cdot \|  \hat q_{1_\textrm{RA}}\|},
\end{equation}
which consists of the projection of the first SPOD mode $\hat q_{1_\textrm{SPOD}}=[\hat u_{1_\textrm{SPOD}}, \hat v_{1_\textrm{SPOD}}]$ on the first resolvent mode $\hat q_{1_\textrm{RA}}=[\hat u_{1_\textrm{RA}}, \hat v_{1_\textrm{RA}}]$. Here $\hat u_1$, $\hat v_1$ are the streamwise and wall-normal components of the first SPOD and first resolvent mode, respectively. Furthermore, $\langle \cdot,\cdot\rangle $ is the $L_2$ inner product and $\|\cdot\|$ is the euclidean norm. The value $\varphi=1$ corresponds to perfect alignment of the modes, whereas $\varphi=0$ indicates that the modes are orthogonal. Note that only the streamwise and wall-normal component are considered here, as the high-speed PIV data  was taken in a monoscopic arrangement for which no information on the spanwise velocity component is available \citep{le2020}.
\section{Results}
\label{sec:results}
\subsection{Global stability analysis}
\begin{figure}
\centering
  \begin{minipage}{0.49\linewidth}
        \includegraphics[trim={0 0 0 0},clip,width=0.95\textwidth]{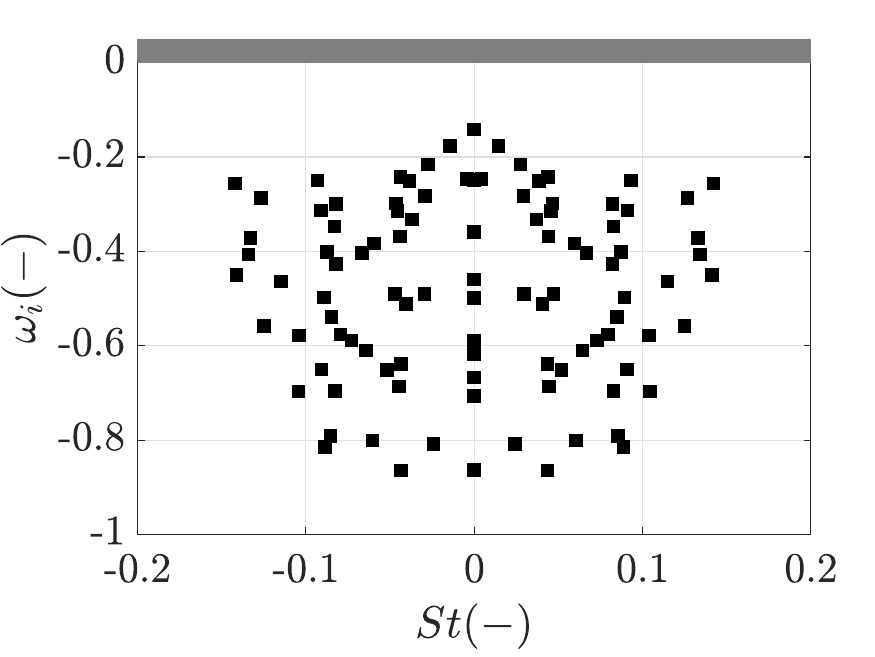}
             \put(-168,130){($a$)}
        \end{minipage}
    \begin{minipage}{0.49\linewidth}
        \centering
         \includegraphics[trim={0 0 0 0},clip,width=0.95\textwidth]{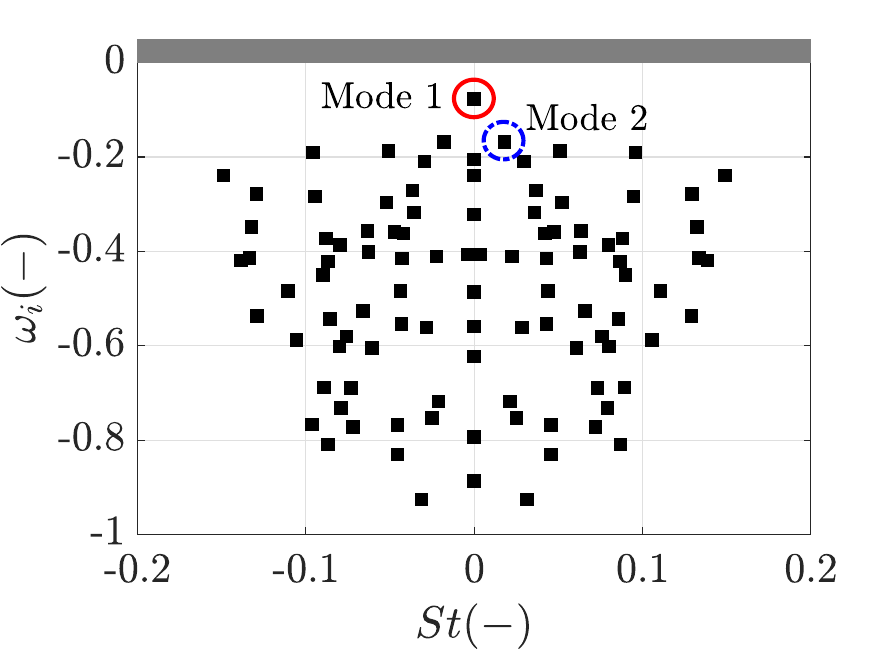}
             \put(-168,130){($b$)}
    \end{minipage}
\caption{Global LSA spectrum for non-dimensional spanwise wavenumber $\beta=0$ ($a$) and $\beta=0.75$ ($b$). The eigenvalue is $\omega=\omega_i+\omega_r$, where $\omega_i$ is the growth rate and the Strouhal number is calculated based on $\omega_r$. The least stable stationary (\solidline) and travelling (\dashedline) mode are highlighted for $\beta=0.75$.}
\label{fig:GLSAspecs}
\end{figure}
We initiate the analysis by investigating the spectral characteristics of the base flow, as well as the influence of the non-dimensional spanwise wavenumber on the GLSA spectrum. Figure \ref{fig:GLSAspecs} illustrates the growth rate of the eigenvalues for the two-dimensional ($\beta=0$) and the three-dimensional case ($\beta=0.75$). Growth rates $\omega_i$ and Strouhal numbers $St=f\cdot L_b/u_\infty$ are depicted. The stability threshold is shown as a grey shaded region. As can be seen, all growth rates $\omega_i$ are negative, indicating that all modes are globally stable.  Moreover, the GLSA spectra display similar shapes, where the least stable mode of each spectrum (2D/3D) is a \say{stationary} mode ($St=0$) and the subsequent modes ($2-4$) are \say{travelling} modes ($St\neq 0$). The stationary nature of the least stable mode of the GLSA is in good agreement with the modal analyses of \cite{theofilis2000}, \cite{robinet2007} and \cite{touber2009} in different types of separation bubbles. In particular, such a stationary global mode was related to a centrifugal instability mechanisms potentially leading to transition in LSBs by \cite{rodriguez2013two}. The present analysis therefore suggests that a similar mechanism might be at play in TSBs. While the latter works revealed the presence of a global instability, the stable nature of the present modes matches well the results of \cite{bugeat2022low} and \cite{sartor2015unsteadiness}, in laminar and turbulent SBLIs, respectively. 

\begin{figure}
\centering
 \includegraphics[trim={0 0 0 0},clip,width=0.85\textwidth]{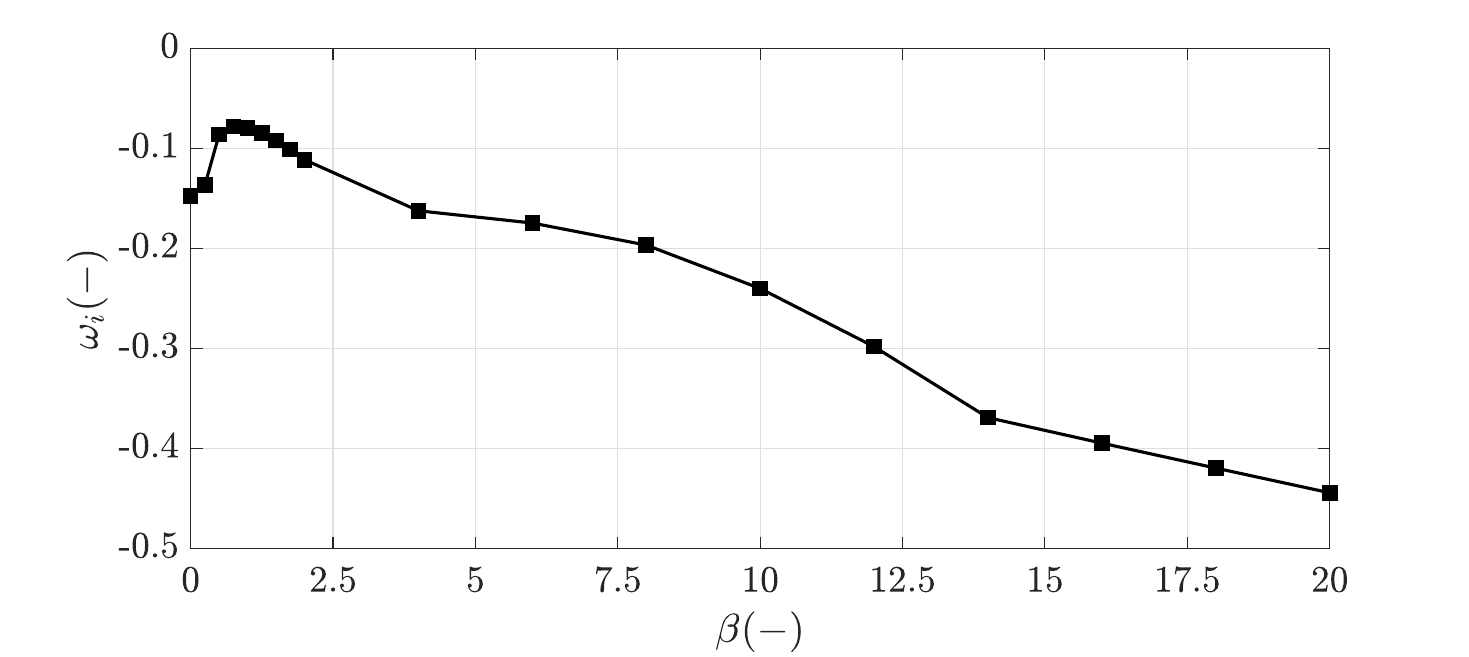}
\caption{Growth rate $\omega_i$ from GLSA over non-dimensional spanwise wavenumber $\beta$. The least damped mode of the spectrum is depicted.}
\label{fig:GLSAmostampl}
\end{figure}

As we progress from two-dimensional ($\beta=0$) towards three-dimensional perturbations ($\beta=0.75$), the least stable mode (Mode 1) moves closer to the stability threshold (figure \ref{fig:GLSAmostampl}). The growth rate $\omega_i$ of the least stable mode reaches its maximum for low non-zero $\beta$, after which it monotonically decreases for all $\beta>0.75$. In relation to the bubble length $L_b$, this particular $\beta$ corresponds to a wavelength $\lambda_z$ approximately eight times the extent of the TSB ($\lambda_z \approx 8 L_b$). This is in good agreement with the low-$\beta$ region observed in the experiments summarized in section \ref{secUnsteady}, where the low-frequency unsteadiness was shown to be characterized by a wavenumber of the order of $\beta \simeq 1$.  Furthermore, the frequency of the least stable travelling mode (Mode 2) takes values between $St=0.014-0.018$ in the low-$\beta$ region ($\beta<2$). This is also consistent with the low-frequency unsteadiness  observed experimentally in the TSB.

\begin{figure}
\centering
\includegraphics[trim={5cm 7.5cm 5cm 2.5cm},clip,width=\textwidth]{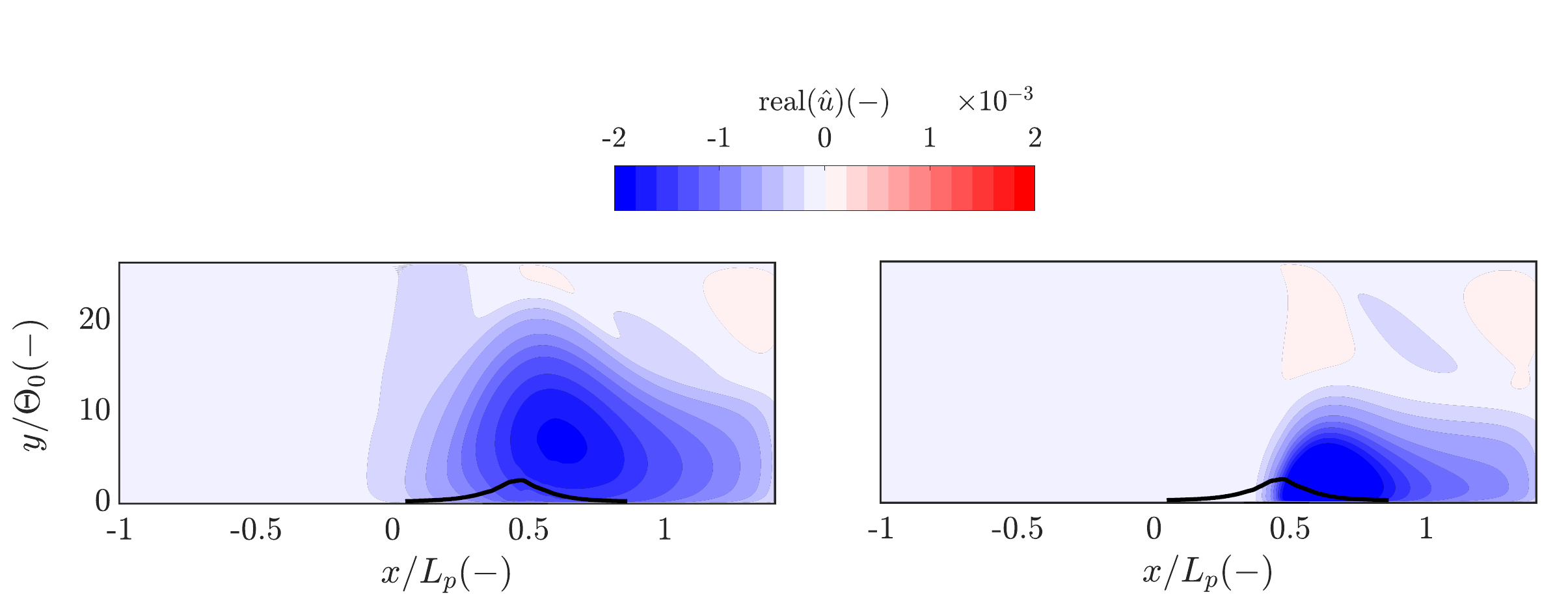}
         \put(-375,83){($a$)}
            \put(-185,83){($b$)}
\caption{Streamwise component of GLSA Mode 1 ($a$) and Mode 2 ($b$) for non-dimensional spanwise wavenumber $\beta=0.75$. The time-averaged location of the TSB is indicated by the dividing streamline $\overline \Psi=0$ (\solidline).}
\label{fig:GLSA_mode12}
\end{figure}

Based on these results, it appears that both the first and second modes in the low-$\beta$ region, although stable, would be reasonable candidates for the low-frequency unsteadiness if properly excited by external perturbations. To further differentiate between modes 1 and 2, we plot their streamwise component in figure \ref{fig:GLSA_mode12}. This particular choice of $\beta$ coincides with the maximum observed in figure \ref{fig:GLSAmostampl}. We observe a least stable stationary mode (Mode 1) that is mainly located in the region surrounding and above the TSB (Fig. \ref{fig:GLSA_mode12} ($a$)). This is consistent with the large-scale, \say{global} nature of the breathing motion, as depicted for example in the SPOD results of figure \ref{fig:SPODmodes}. While the corresponding travelling mode (Mode 2) shares qualitative similarities with Mode 1, we note that the observed structure is shifted towards the downstream part of the bubble and smaller in the wall-normal and streamwise directions (Fig. \ref{fig:GLSA_mode12} ($b$)). Hence, based on its shape, relatively small decay rate at low spanwise wavenumber, and stationary character, Mode 1 thus appears to be a better candidate for the low-frequency unsteadiness. However, its stable character indicates that this mode is not capable of generating self-sustained oscillations in the TSB. We now consider resolvent analysis to investigate the receptivity of this mode to external perturbations.
\subsection{Resolvent analysis}
\begin{figure}
\centering
\includegraphics[trim={0 0 0 0},clip,width=0.85\textwidth]{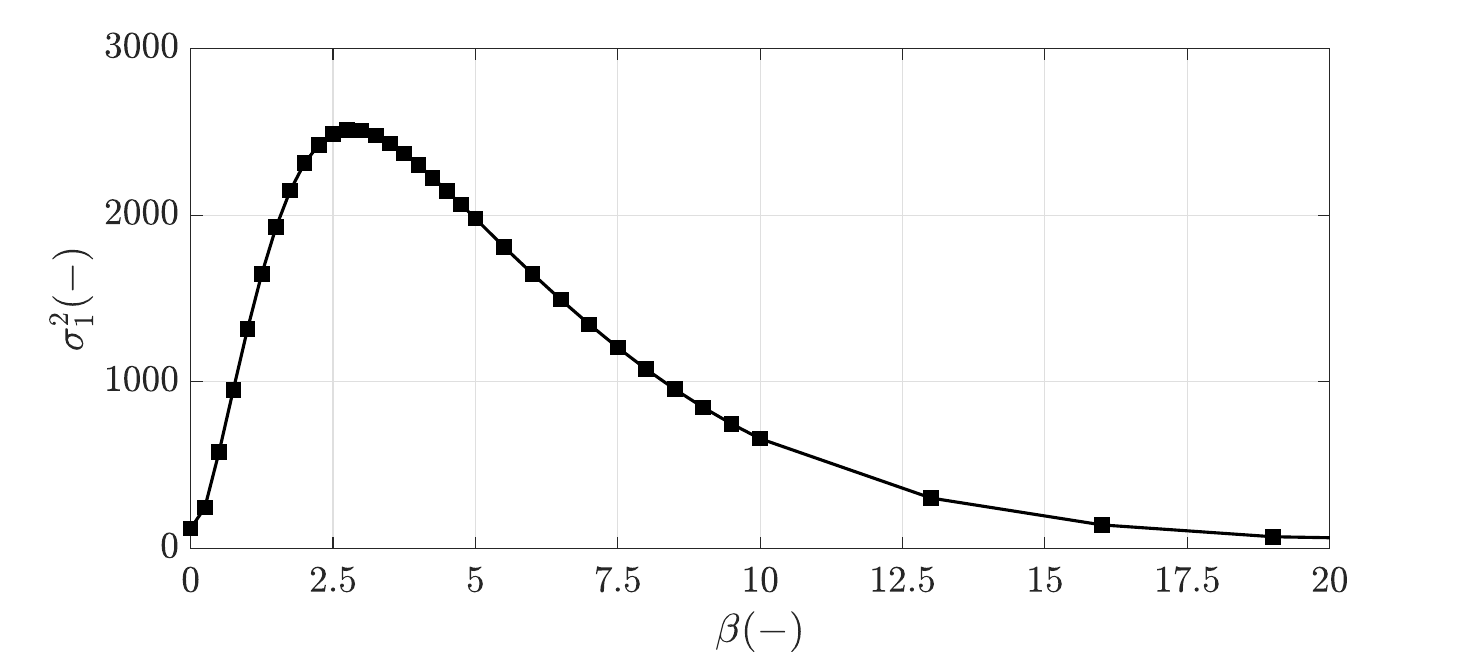}
\caption{Optimal energy gain $\sigma_1^2$ from resolvent analysis for different non-dimensional spanwise wavenumbers $\beta$. The frequency is fixed at $St=0.01$.}
\label{fig:RA_sigmavsbeta}
\end{figure}

In the previous section we established that the investigated TSB flow is globally stable in the asymptotic time limit. In order to study its response to forcing, we now proceed to outline results from the resolvent analysis, directing particular focus towards the low-$\beta$ region as well as the low-frequency range. In figure \ref{fig:RA_sigmavsbeta} the optimal energy gain $\sigma_1^2$ of the RA is shown for different non-dimensional spanwise wavenumbers $\beta$. The Strouhal number is fixed at $St=0.01$, corresponding to the characteristic low-frequency breathing motion. The optimal energy gain increases up to a maximum of $\beta=2.75$, after which it monotonically decreases for  $\beta>2.75$. The peak amplitude is thus reached at a non-dimensional spanwise wavenumber that is slightly higher that for GLSA. The wavenumber $\beta=2.75$ is also somewhat larger than the low-$\beta$ region defined in section \ref{secUnsteady}. However, in relation to the bubble length $L_b$, this particular $\beta$ corresponds to a wavelength $\lambda_z$ approximately twice the extent of the TSB ($\lambda_z \approx 2 L_b$). We therefore conclude that the value $\beta=2.75$ can still be considered low-$\beta$, as the associated wavelength is substantially larger than the TSB. In order to allow for a consistent comparison between GLSA and RA results, we will proceed to further analyze the non-dimensional spanwise wavenumbers $\beta=0$ and $\beta=0.75$. 

\begin{figure}
\centering
     \begin{minipage}{0.49\linewidth}
                     \includegraphics[trim={0 0 0 0},clip,width=0.95\textwidth]{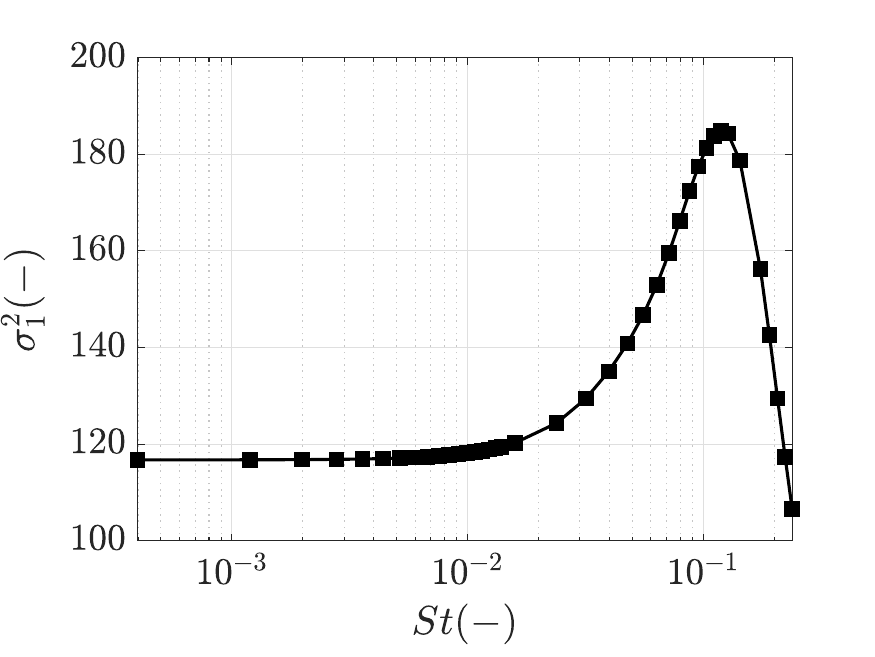}
        \put(-180,125){($a$)}
    \end{minipage}
    \begin{minipage}{0.49\linewidth}
        \centering
         \includegraphics[trim={0 0 0 0},clip,width=0.95\textwidth]{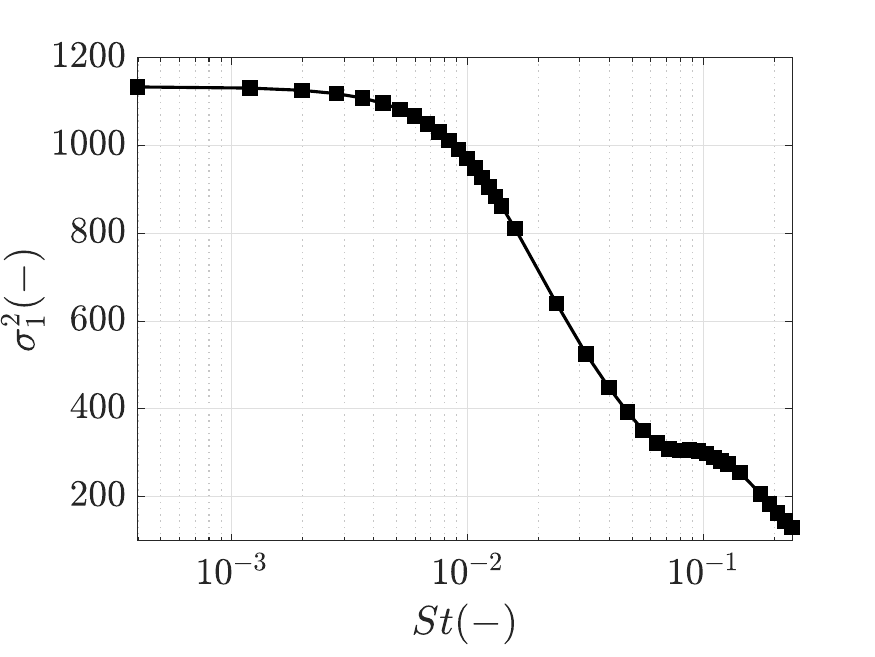}
              \put(-180,125){($b$)}
    \end{minipage}
\caption{Optimal energy gain $\sigma_1^2$ from resolvent analysis versus different Strouhal numbers $St$ for two-dimensional case $\beta=0$ ($a$) and a sample three-dimensional case $\beta=0.75$ ($b$).}
\label{fig:RA_spectra_2D_3D}
\end{figure}

\begin{figure}
\centering
 \includegraphics[trim={5cm 7.5cm 5cm 2.5cm},clip,width=\textwidth]{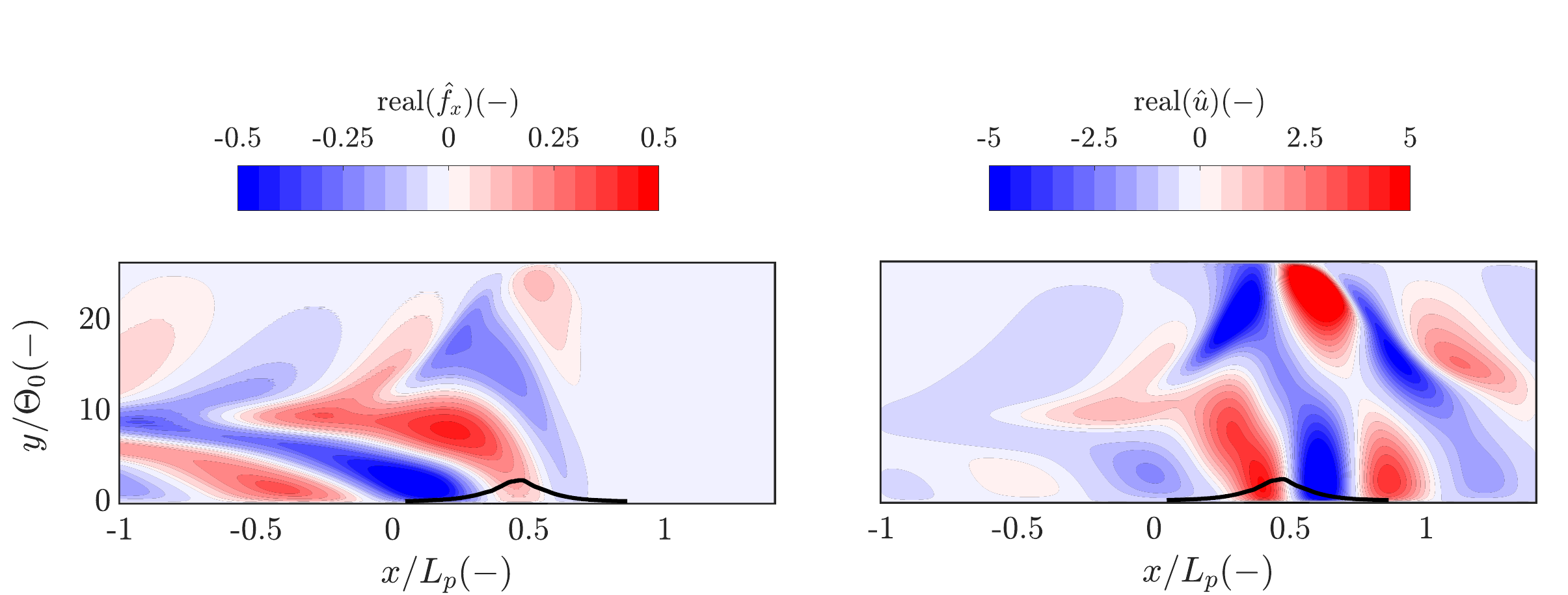}
         \put(-375,83){($a$)}
            \put(-185,83){($b$)}
\caption{Streamwise component of optimal forcing $\hat f_x$  ($a$) and response $\hat u$ ($b$) for non-dimensional spanwise wavenumber $\beta=0$ and $St=0.12$. The time-averaged location of the TSB is indicated by the dividing streamline $\overline \Psi=0$ (\solidline).}
\label{fig:RA_shedding}
\end{figure}

We now investigate the behavior of the optimal energy gain $\sigma_1^2$ for different Strouhal numbers $St$, where we distinguish between the two-dimensional case ($\beta=0$) and the three-dimensional case ($\beta=0.75$). For $\beta=0$, we obtain a distribution that monotonically increases up to a Strouhal number of $St\approx 0.1$, after which the optimal energy gain decreases again (Fig. \ref{fig:RA_spectra_2D_3D} ($a$)). The associated (streamwise) optimal forcing and response are displayed in figure \ref{fig:RA_shedding}. At medium frequency ($St\approx 0.1$) and for two-dimensional perturbations ($\beta=0$), the optimal forcing is located mostly upstream of the bubble. The resulting optimal response exhibits the typical alternating pattern of the Kelvin-Helmholtz (K-H) instability waves, which are tilted towards the direction of the base flow shear. For the present DNS base flow, it was shown in \cite{cura2023medium} by means of local linear stability analysis, that the medium-frequency regime, characterized by the Strouhal number $St\approx 0.1$, is related to the roll-up and shedding of vortices through a shear layer instability. The present results confirm the latter analysis. Furthermore, the revealed structures strongly resemble the medium-frequency modes for the turbulent SWBLI from \cite{sartor2015unsteadiness} and the laminar SWBLI from \cite{bugeat2022low}. However, in the latter they were attributed to a convective instability reminiscent of the compressible counterpart of the Tollmien-Schlichting instability.

In the case of three-dimensional perturbations ($\beta=0.75$), a substantial difference can be observed (Fig. \ref{fig:RA_spectra_2D_3D} ($b$)). The highest energy gains are now obtained in the low-frequency range, after which a distribution similar to that of a low-pass filter becomes apparent. Here, a distinct drop in optimal gain $\sigma_1^2$ occurs in the range $St=10^{-3}-10^{-2}$, which is in good agreement with the Strouhal number $St\simeq0.01$ reported in pre-multiplied PSDs describing the low-frequency unsteadiness (e.g., Fig. \ref{pic:Pressure}). We can further observe a small \say{hump} in the region of $St\approx 0.1$, which, yet again, can be associated with K-H instability. Interestingly, this medium-frequency hump is only observed in the narrow range of non-dimensional spanwise wavenumbers $0<\beta<1.25$. It is most pronounced for $\beta=0.25$ and disappears for $\beta>1$. The low-pass filter behavior on the other hand, can be observed for any low non-zero $\beta$.

\begin{figure}
\centering
     \begin{minipage}{0.49\linewidth}
               \includegraphics[trim={0 0 0 0},clip,width=0.95\textwidth]{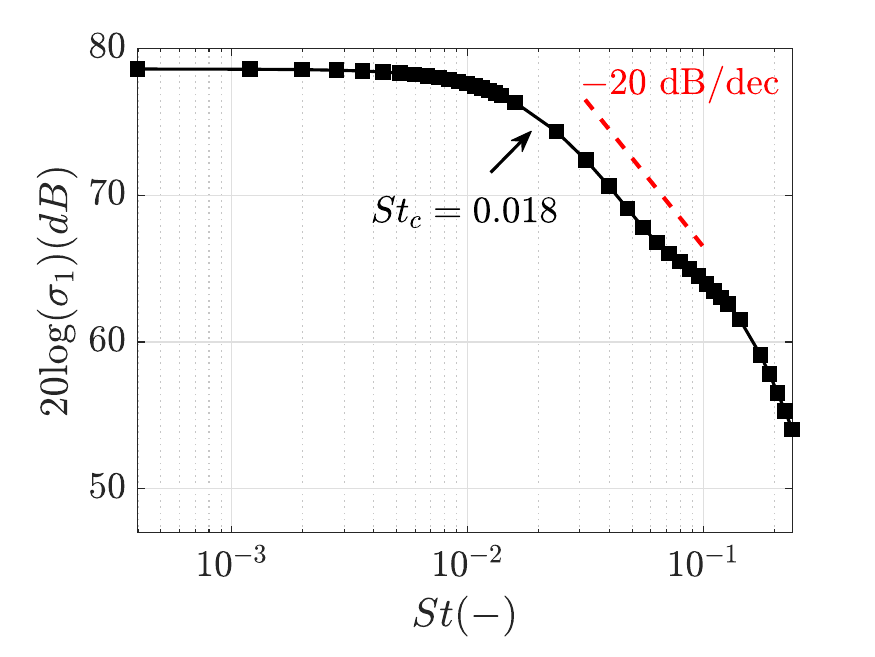}
               \put(-180,125){($a$)}
    \end{minipage}
    \begin{minipage}{0.49\linewidth}
        \centering
                \includegraphics[trim={0 0 0 0},clip,width=0.95\textwidth]{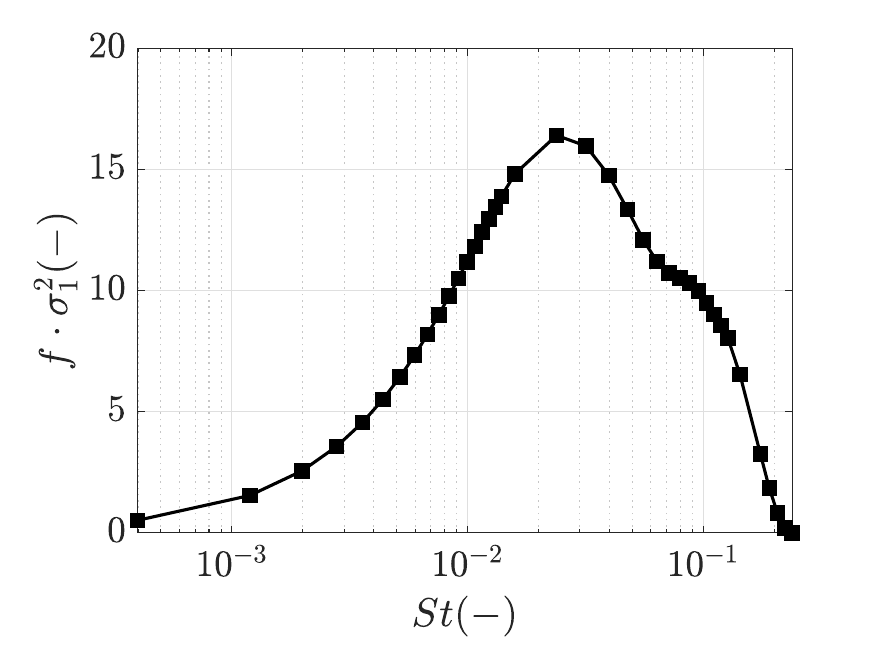}
                \put(-180,125){($b$)}
    \end{minipage}
\caption{First-order low-pass filter representation of optimal gain from resolvent analysis $20\textrm{log}(\sigma_1)$ ($a$) and pre-multiplied energy gain $f\cdot \sigma_1^2$ ($b$) versus different Strouhal numbers $St$ ($b$). The non-dimensional spanwise wavenumber is $\beta=2$. The Strouhal number $St_c$ corresponding to the cut-off frequency based on the $-\SI{3}{\decibel}$ rule is indicated in the figure. We further depict the characteristic slope of a first-order low-pass filter $-\SI{20}{\decibel}/\textrm{dec}$ (\dashedline).}
\label{fig:RA_sigmaDB}
\end{figure}

We now define a decibel scale for the optimal gain $\sigma_1$
\begin{equation}
    \sigma_{1,\textrm{dB}}(St)=20\textrm{log}(\sigma_1(St))
\end{equation}
and apply the $-\SI{3}{\decibel}$ rule to estimate the cut-off frequency of the filter function. This is shown in figure \ref{fig:RA_sigmaDB} ($a$), where the low-pass filter representation of $\sigma_1$ is shown for the non-dimensional spanwise wavenumber $\beta=2$. We obtain a cut-off frequency corresponding to $St_c=0.018$. Interestingly, the cut-off frequency monotonically increases from $\beta=0.25$ to $\beta=2$ and takes values between $0.006$ and $0.018$. Hence, all non-dimensional spanwise wavenumbers associated with the low non-zero-$\beta$ regime reveal cut-off frequencies which are in good agreement with values reported for the low-frequency breathing phenomenon $St\simeq0.01$ (e.g. \cite{mohammed2016, le2020, steinfurth2022, richardson2023}). Furthermore, from figure \ref{fig:RA_sigmaDB} ($a$) it becomes apparent that the low-pass filter representation of $\sigma_1$ indicates a first-order behaviour. After the cut-off frequency $St_c$, the amplitude of $20\textrm{log}(\sigma_1)$ decreases approximately at the rate $-20\SI{}{\decibel}/\textrm{dec}$, as indicated by the red-dashed line in figure \ref{fig:RA_sigmaDB} ($a$). This behavior can be observed for any low non-zero $\beta$. Interestingly, this first-order-filter character of low-frequency unsteadiness has already been observed in the incompressible TSB of \cite{mohammed2021}. It is also a recurrent observation in turbulent SBLIs \citep{Plotkin1975,Touber2011,poggie2015}. Furthermore, a similar low-pass filter model has also been found in the resolvent analysis of a laminar SBLI by \cite{bugeat2022low}. Combined with these existing results, our new findings suggest that the low-pass filter model of separation bubble unsteadiness may be valid in a broad range of Reynolds and Mach numbers. In pre-multiplied form, the frequency response of first-order filters typically shows a maximum at a frequency close to the $-3$ dB cut off \citep{poggie2015}. As can be observed in figure \ref{fig:RA_sigmaDB} ($b$), such a hump can also be observed for the energy gain  $f\cdot \sigma_1^2$. The resulting curve is also similar to the pre-multiplied PSD of the wall pressure fluctuations shown in figure \ref{pic:Pressure} ($b$).

\begin{figure}
\centering
\includegraphics[trim={5cm 7.5cm 5cm 2.5cm},clip,width=\textwidth]{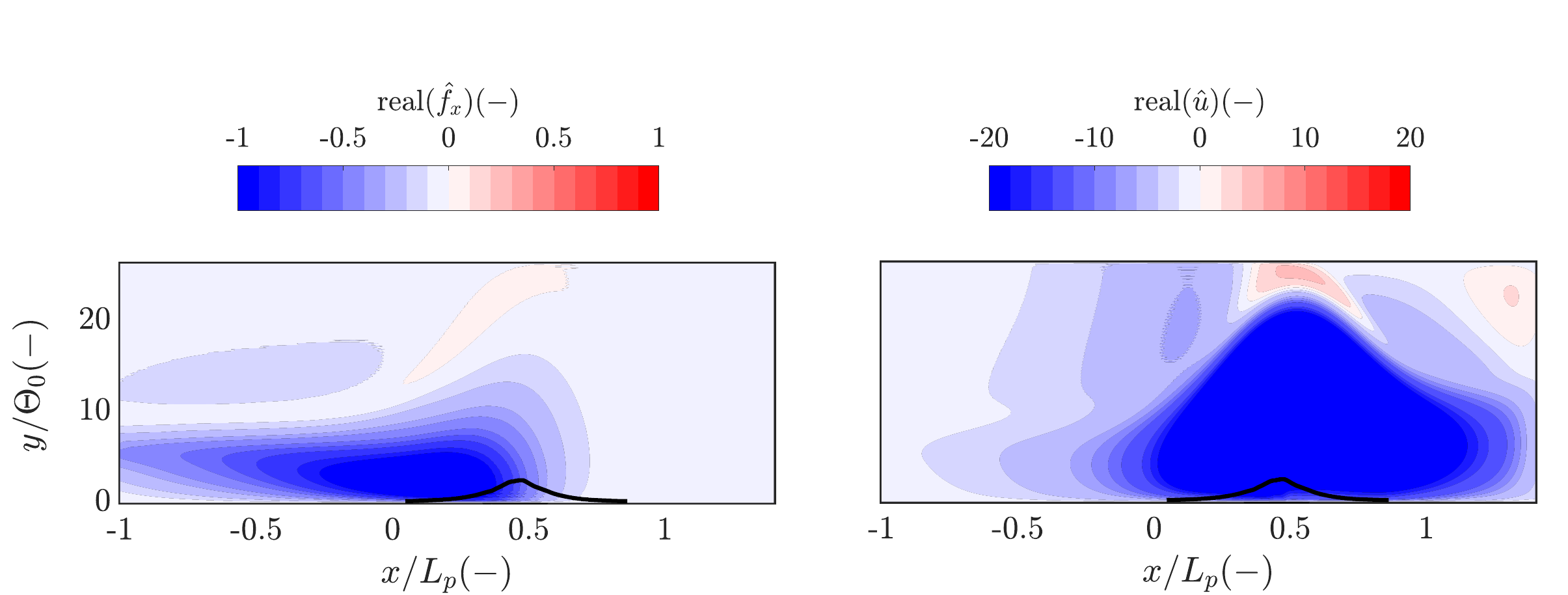}
         \put(-375,83){($a$)}
            \put(-185,83){($b$)}
\caption{Streamwise component of optimal forcing $\hat f_x$ ($a$) and response $\hat u$ ($b$) for non-dimensional spanwise wavenumber $\beta=0.75$ and $St=0.01$. The time-averaged location of the TSB is indicated by the dividing streamline $\overline \Psi=0$ (\solidline).}
\label{fig:RA_breathing}
\end{figure}
In figure \ref{fig:RA_breathing} the (streamwise) optimal forcing and response associated with the low-frequency, low non-zero-$\beta$ regime are displayed. This is exemplary shown for $\beta=0.75$ and $St=0.01$. The optimal forcing is located mostly upstream of the bubble, encompassing a large portion of the domain. A large elongated structure can be observed, where the maximum amplitude is reached near the time-averaged separation point (Fig. \ref{fig:RA_breathing} ($a$)). The resulting optimal response bounds the recirculation region and follows its shape (Fig. \ref{fig:RA_breathing} ($b$)). Similar to the low-pass filter behavior of the optimal energy gain (Fig. \ref{fig:RA_spectra_2D_3D}), any non-dimensional spanwise wavenumber in the low non-zero-$\beta$ regime exhibits similar low-frequency characteristics. That is, similar contours of optimal forcing and response could be observed for any ($St, \beta$) within this region. This behavior suggests that the receptivity at low-frequency is not driven by a unique frequency and spanwise wavenumber, but rather by a range of low Strouhal numbers ($St\simeq 0.01$) and low non-zero spanwise wavenumbers. Crucially, the shape and position of the optimal response closely resemble the least stable mode (Mode 1) of the GLSA. Therefore, our results suggest that the low-frequency unsteadiness in our TSB is driven by a modal mechanism, where the least stable global mode is selectively amplified for a range of ($St, \beta$). In turn, since this global mode may be associated with a stable centrifugal mechanism, the results suggest that the TSB low-frequency unsteadiness is related to the excitation of such centrifugal dynamics by turbulence. This hypothesis will be further discussed in section \ref{sec:disc}.
\subsubsection{Low non-zero-$\beta$ region}
\begin{figure}
\centering
     \begin{minipage}{0.49\linewidth}
        \includegraphics[trim={0 0 0 0},clip,width=0.95\textwidth]{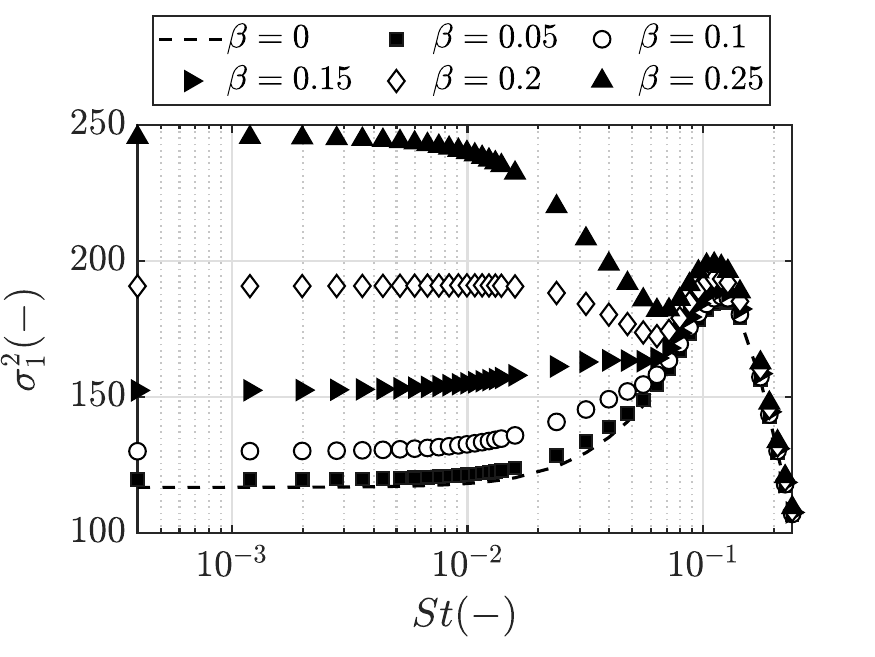}
    \end{minipage}
    \caption{Optimal energy gain $\sigma_1^2$ from resolvent analysis versus different Strouhal numbers $St$ for small spanwise wavenumbers $\beta\le0.25$. The represented optimal energy gains correspond to $\beta=0$ (\dashedline), $\beta=0.05$ (\squaresymb), $\beta=0.1$ (\hollowcircle), $\beta=0.15$ (\solidlefttriangle), $\beta=0.2$ (\hollowdiamond) and $\beta=0.2$ (\solidtriangle).}
    \label{pic:2Dto3D_RA}
\end{figure}

So far, we have used the term \say{low non-zero spanwise wavenumber} loosely, assuming that certain properties of the optimal gain as well as their associated forcing and response can be recovered as long as $\beta \neq 0$ and $\lambda_z\gg L_b$. However, the transition from what we observe for two-dimensional perturbations ($\beta=0$) to the behavior of what we dubbed the low non-zero-$\beta$ region is not discrete, but rather continuous. Hence, for the sake of clarity and completeness, in this section, we take a closer look at the region $\beta\le0.25$. In figure \ref{pic:2Dto3D_RA} we plot the optimal energy gain $\sigma_1^2$ over different Strouhal numbers $St$ for six spanwise wavenumbers $\beta \in [0, 0.25]$. For $\beta=0$, we obtain the same distribution as shown in figure \ref{fig:RA_spectra_2D_3D} ($a$). The distinct peak at $St\simeq 0.1$, which we previously associated with the K-H instability, remains visible for all $\beta$ shown in figure \ref{pic:2Dto3D_RA}. However, as $\beta$ increases, the optimal energy gain $\sigma_1^2$ in the low-frequency region increases as well. Finally, for $\beta=0.25$ we first observe the expected low-pass filter behavior, with dominant optimal energy gains in the low-frequency region. The associated contours of optimal forcing and response now appear similar to figure \ref{fig:RA_breathing} (not shown here). This behavior can be observed up to approximately $\beta=3$, after which the qualitative shape of optimal forcing and response starts to differ. Taking this into account, we may now adopt the (loose) definition of $\beta \in [0.25, 3]$ whenever we refer to the low non-zero-$\beta$ region.
\subsubsection{Three-dimensional structures}
\begin{figure}
\centering
    \begin{minipage}{0.49\linewidth}
                \includegraphics[trim={1.5cm 8cm 1.5cm 10cm},clip,width=0.95\textwidth]{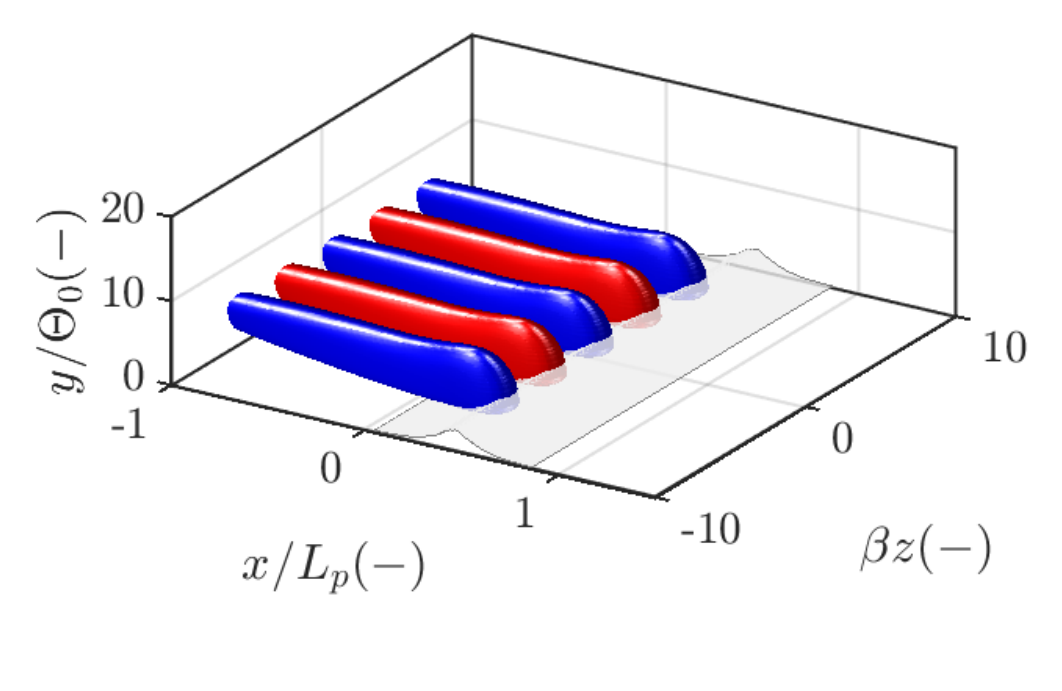}
         \put(-170,100){($a$)}
    \end{minipage}
    \begin{minipage}{0.49\linewidth}
        \centering
                  \includegraphics[trim={1.5cm 8cm 1.5cm 10cm},clip,width=0.95\textwidth]{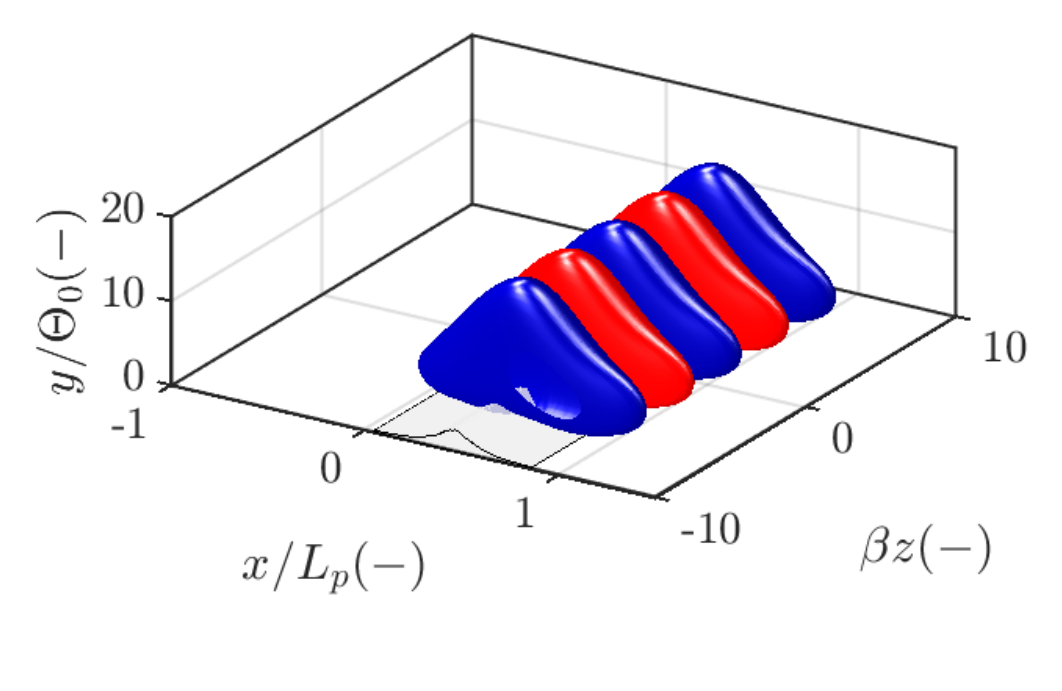}
         \put(-170,100){($b$)}
    \end{minipage}
    \begin{minipage}{0.49\linewidth}
                \includegraphics[trim={1.5cm 8cm 1.5cm 10cm},clip,width=0.95\textwidth]{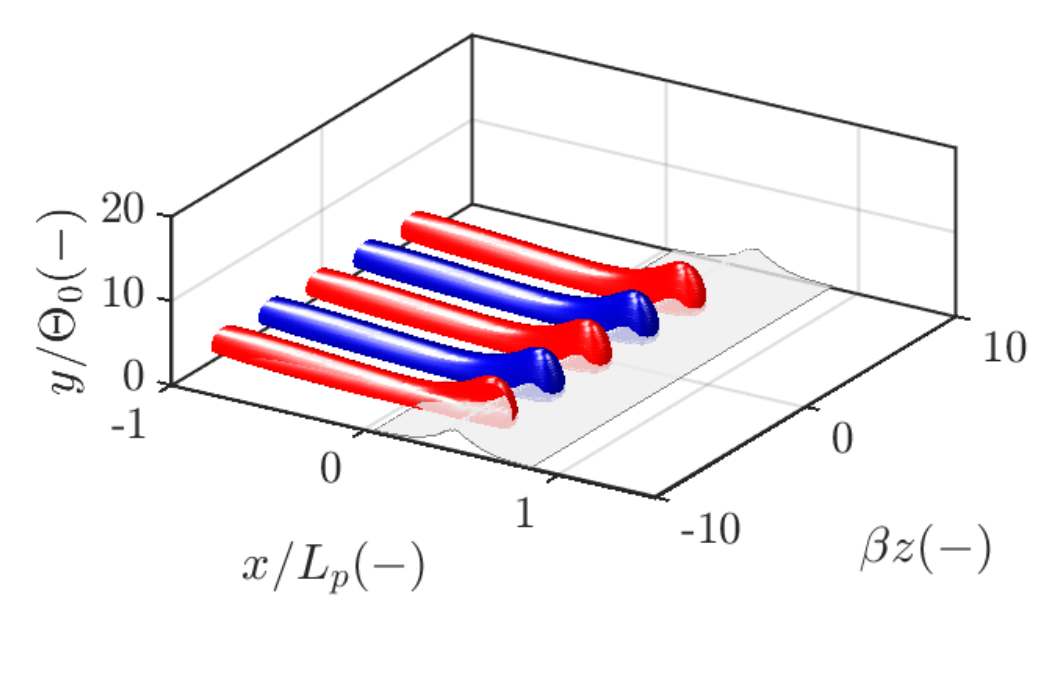}
        \put(-170,100){($c$)}
    \end{minipage}
    \begin{minipage}{0.49\linewidth}
        \centering
                 \includegraphics[trim={1.5cm 8cm 1.5cm 10cm},clip,width=0.95\textwidth]{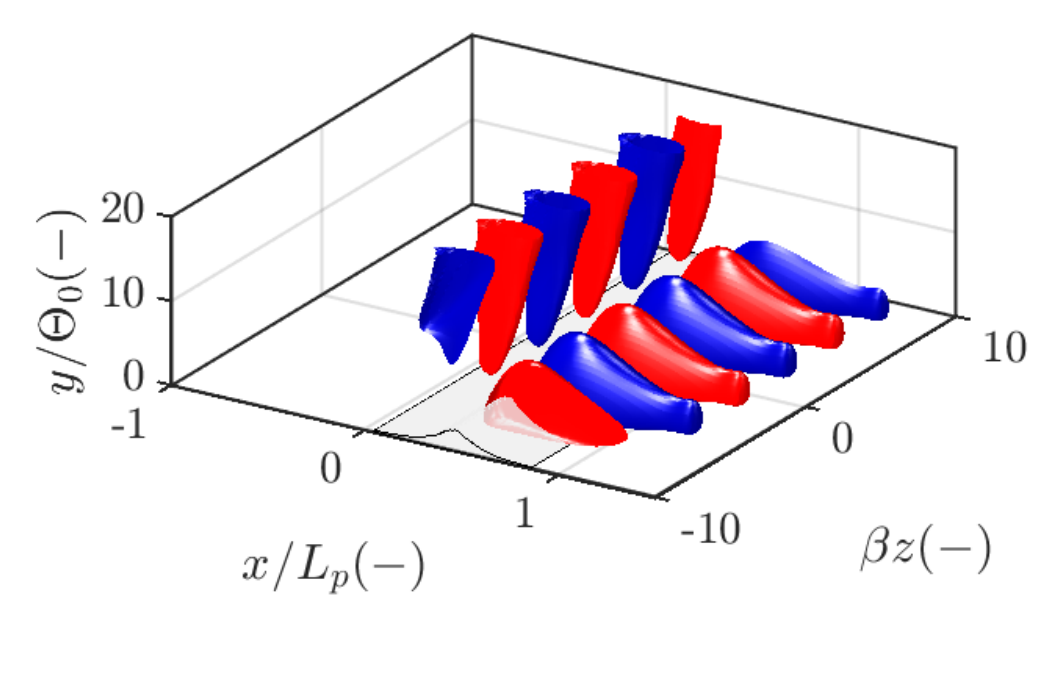}
         \put(-170,100){($d$)}
    \end{minipage}
\caption{Streamwise component of optimal forcing $ \hat f_x$ ($a$) and response $ \hat u$ ($b$) and spanwise component of optimal forcing $ \hat f_z$ ($c$) and response $ \hat w$ ($d$) from RA for non-dimensional spanwise wavenumber $\beta=0.75$. The real part of the modes is depicted. Iso-surfaces of $\pm \SI{35}{\percent}$ of $\textrm{max}|\hat f_x|$, $\textrm{max}|\hat u|$ and $\textrm{max}|\hat f_z|$, $\textrm{max}|\hat w|$ are shown, respectively.}
\label{fig:3D}
\end{figure}

In the following, we investigate the three-dimensional structure of the resolvent modes in the low non-zero-$\beta$, low-frequency regime. For this purpose, we plot iso-surfaces of the streamwise (Fig. \ref{fig:3D} ($a,b$)) and spanwise (Fig. \ref{fig:3D} ($c,d$)) component of the optimal forcing and response. Iso-surfaces of $\pm \SI{35}{\percent}$ of the forcings $|\hat f_x|, |\hat f_z|$ and the responses $|\hat u|, |\hat w|$ are displayed, respectively. The time-averaged location of the separation bubble is indicated by the grey-shaded region, which corresponds to the dividing streamline ($\overline \Psi=0$). The frequency is fixed at $St=0.01$ and the non-dimensional spanwise wavenumber is $\beta=0.75$. In this representation, the distinction between optimal forcing and response is even more prominent. While the optimal forcing is located mostly upstream and in the first half of the TSB (Figs. \ref{fig:3D} ($a,c$)), the optimal responses $\hat u, \hat w$ are located in the region surrounding the separation bubble. This particular spatial configuration of forcing and response suggests that upstream disturbances are a required contribution to the low-frequency unsteadiness. Notably, the elongated structures in the upstream region closely resemble boundary layer streaks, albeit with much larger spanwise wavelength compared to usual superstructures in turbulent boundary layers, as discussed in section \ref{sec:expdata}. Whereas the streamwise component of forcing is concentrated to areas parallel to the bottom wall (Fig. \ref{fig:3D} ($a$)), the structure of the spanwise component is slightly tilted upward in the TSB region in figure \ref{fig:3D} ($c$). This behavior was also observed for the spanwise forcing at low $\beta$ in \cite{bugeat2022low}, even though the investigated flow was a laminar SBLI.

\subsubsection{Sub-optimal energy gains}
\begin{figure}
\centering
 \includegraphics[trim={0 0 0 0},clip,width=0.85\textwidth]{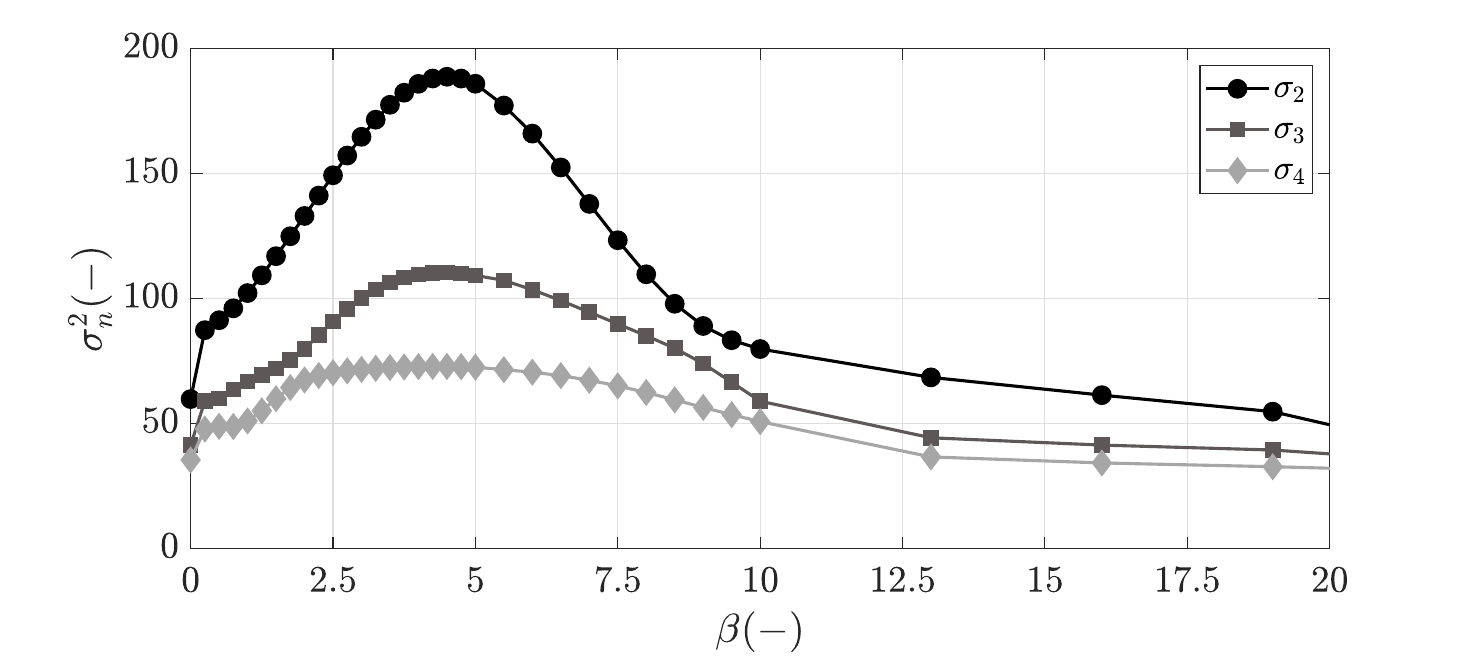}
\caption{Sub-optimal energy gain $\sigma_n^2$ from resolvent analysis for different non-dimensional spanwise wavenumbers $\beta$. The frequency is fixed at $St=0.01$. The displayed sub-optimal gains are $\sigma_2$ (\solidlinesolidcirc), $\sigma_3$ (\solidlinesquare) and $\sigma_4$ (\solidlinediamond).}
\label{fig:RA_sigmavsbeta_subopt}
\end{figure}

\begin{figure}
\centering
     \begin{minipage}{0.49\linewidth}
     \includegraphics[trim={0 0 0 0},clip,width=0.95\textwidth]{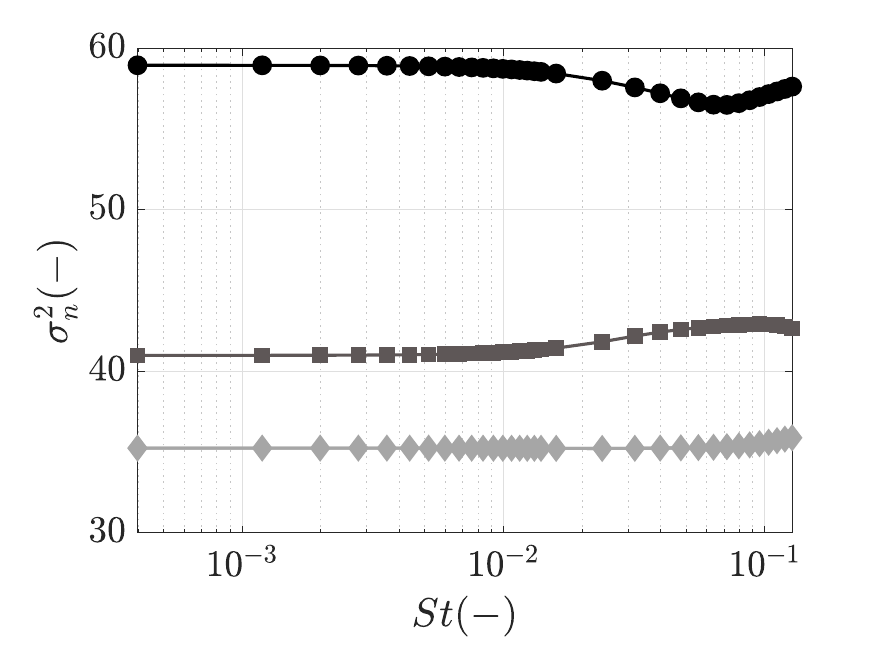}
               \put(-180,125){($a$)}
    \end{minipage}
    \begin{minipage}{0.49\linewidth}
        \centering
         \includegraphics[trim={0 0 0 0},clip,width=0.95\textwidth]{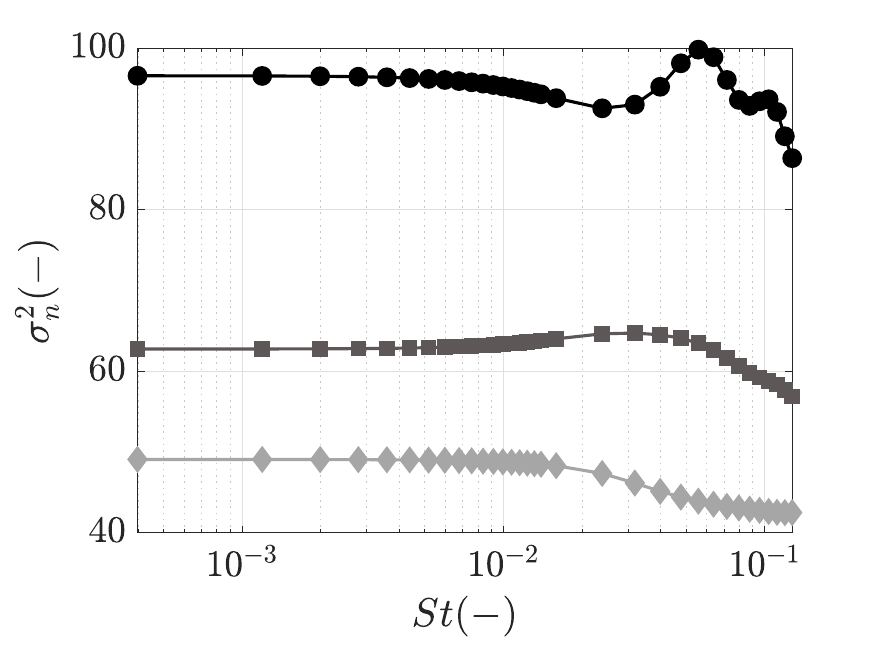}
                \put(-180,125){($b$)}
    \end{minipage}
\caption{Sub-optimal energy gains $\sigma_n^2$ from resolvent analysis versus different Strouhal numbers $St$ for two-dimensional case $\beta=0$ ($a$) and a sample three-dimensional case $\beta=0.75$ ($b$). The displayed sub-optimal energy gains are $\sigma_2^2$ (\solidlinesolidcirc), $\sigma_3^2$ (\solidlinesquare) and $\sigma_4^2$ (\solidlinediamond)}
\label{fig:RA_spectra_2D_3D_subopt}
\end{figure}

In this section, the behavior of the sub-optimal resolvent modes associated with $\sigma_2, \sigma_3, ..., \sigma_n$ and their corresponding forcing fields is studied. Figure \ref{fig:RA_sigmavsbeta_subopt} shows the sub-optimal energy gains $\sigma_n^2$ at the fixed frequency $St=0.01$ and for different non-dimensional spanwise wavenumbers $\beta$. Similar trends as for the optimal energy gain $\sigma_1^2$ can be observed. The sub-optimal energy gains monotonically increase up to a maximum at low/moderate $\beta$, after which they start to decay. The maximum of the curves, however, is slightly shifted towards higher $\beta$ as compared to figure \ref{fig:RA_sigmavsbeta}. For instance, $\sigma_2^2$ attains its maximum value for $\beta=4.5$, which no longer aligns well with the previously defined region of low non-zero $\beta$ observed in the experiments. Furthermore, the low-pass filter behavior observed in the optimal energy gain $\sigma_1^2$ is not as pronounced for the sub-optimal energy gains. Their behavior over different Strouhal numbers is shown in figure \ref{fig:RA_spectra_2D_3D_subopt} for the two-dimensional case ($\beta=0$) and the three-dimensional case ($\beta=0.75$). For three-dimensional perturbations ($\beta=0.75$) the sub-optimal energy gain $\sigma_2^2$ exhibits a very strong hump at $St=0.06$ and a secondary hump at $St=0.1$ (Fig. \ref{fig:RA_spectra_2D_3D_subopt} ($b$)). The low-pass filter behavior is only visible for $St<0.02$. Similar observations can be made for $\sigma_3^2$. Here, the low-pass filter function is superimposed by a broadband hump centered around $St=0.03$. Only the sub-optimal energy gain $\sigma_4^2$ displays the expected low-pass filter behavior with a reasonable cut-off frequency.

\begin{figure}
\centering
 \includegraphics[trim={5cm 7.5cm 5cm 2.5cm},clip,width=\textwidth]{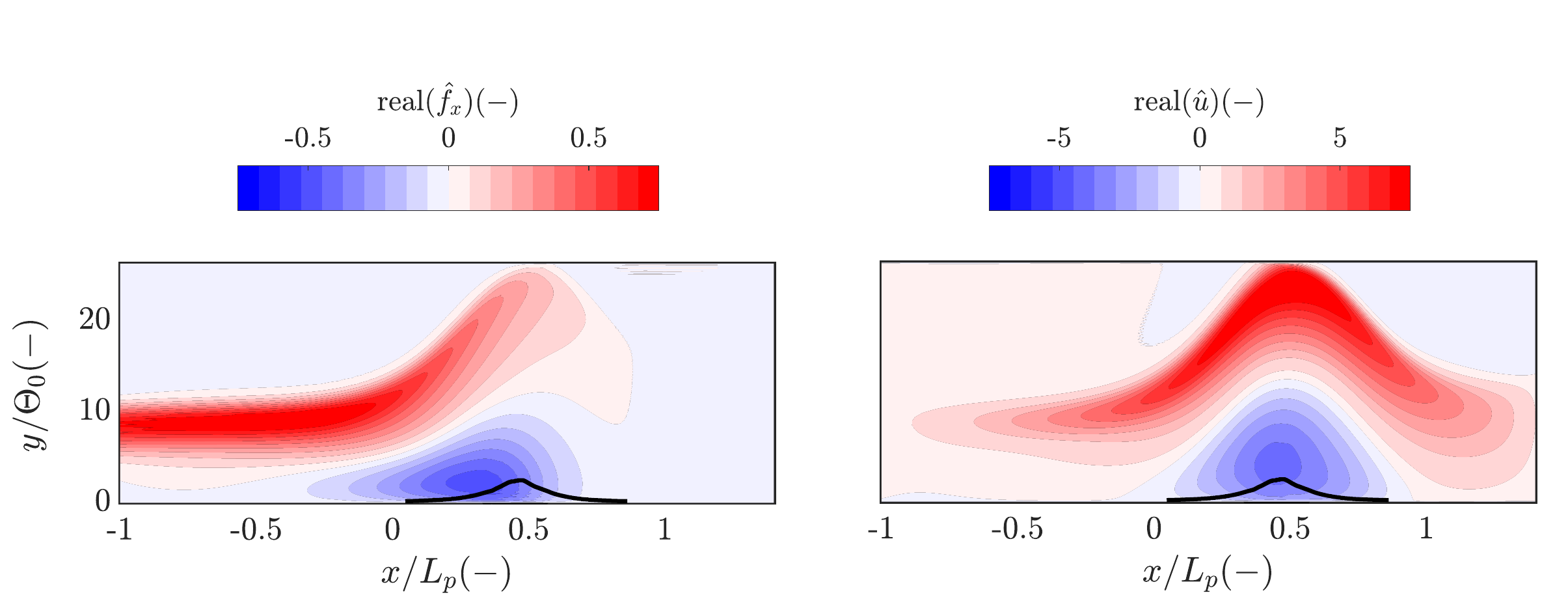}
          \put(-375,83){($a$)}
            \put(-185,83){($b$)}
\caption{Streamwise component of sub-optimal forcing $\hat f_x$ ($a$) and response $\hat u$ ($b$) related to the sub-optimal energy gain $\sigma_2^2$. The non-dimensional spanwise wavenumber amd frequency are $\beta=0.75$ and $St=0.01$, respectively. The time-averaged location of the TSB is indicated by the dividing streamline $\overline \Psi=0$ (\solidline).}
\label{fig:RA_subopt2}
\end{figure}

Generally speaking, the sub-optimal forcings and responses retain similar features as the optimal forcings and responses (figures \ref{fig:RA_shedding}, \ref{fig:RA_breathing}). However, additional structures away from the bubble and oriented along the base flow shear become apparent, see for example figure \ref{fig:RA_subopt2}. Combined with the fact that the optimal gain is low rank, as indicated by $\sigma_1^2/\sigma_2^2 \approx 12$, this leads to the inference that the sub-optimal gains are less likely to be responsible for the receptivity at low frequency.

\subsubsection{Effect of eddy viscosity}

\begin{figure}
\centering
 \includegraphics[trim={0 0 0 0},clip,width=0.85\textwidth]{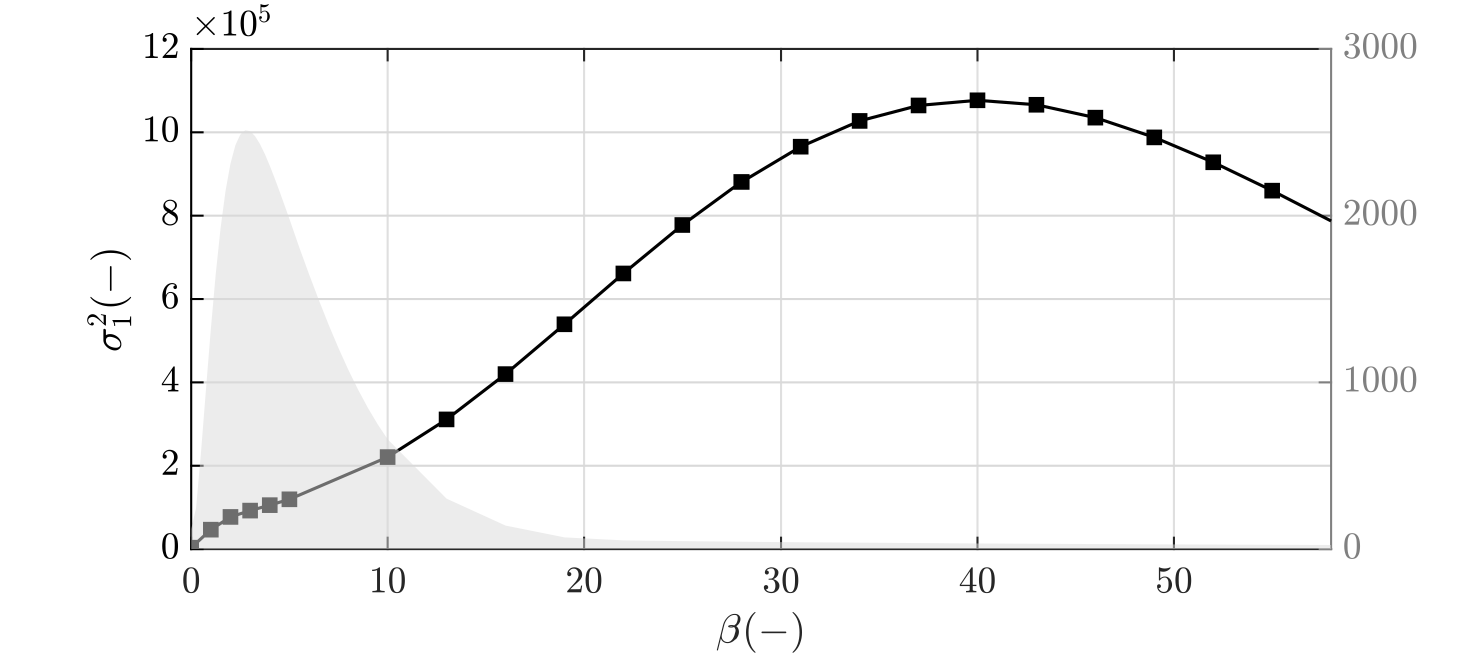}
\caption{Optimal energy gain $\sigma_1^2$ from resolvent analysis when the terms related to the eddy-viscosity are neglected ($\nu_t=0$) for different non-dimensional spanwise wavenumbers $\beta$ (\solidlinesquare). The frequency is fixed at $St=0.01$. The optimal energy gain including the eddy-viscosity model in the LNSE operator is indicated as a grey-shaded region.}
\label{fig:RA_nut_0}
\end{figure}

Up to this point, all presented results have been computed using the LNSE operator modelling part of the Reynolds stresses by means of an eddy-viscosity model. Here we briefly discuss the effects of neglecting the eddy viscosity $(\nu_t=0)$ on the forced dynamics of the bubble. In figure  \ref{fig:RA_nut_0} the optimal energy gain $\sigma_1^2$ is displayed for the fixed frequency $St=0.01$ and over different non-dimensional spanwise wavenumbers $\beta$. A strongly modified curve can be observed as compared to figure \ref{fig:RA_sigmavsbeta}, where $\nu_t \neq 0$ (grey-shaded region). The distinct peak observed for $\beta=2.75$ is now barely visible. Only a very slight \say{plateau} in the expected region of spanwise wavenumbers remains. On the other hand, a maximum for high non-dimensional spanwise wavenumbers becomes apparent at $\beta=40$ ($\lambda_z\approx 0.15 L_b$). This curve looks very similar to the distribution recovered in \cite{bugeat2022low} in a laminar SBLI, nevertheless, the local maximum at low non-zero $\beta$ is less distinct in the present work. Remarkably, the modes at low non-zero $\beta$ and low frequency $St\approx 0.01$ (not shown here) look very similar, regardless of whether or not a $\nu_t$-model is employed.

\section{Discussion}
\label{sec:disc}

The results presented in section \ref{sec:results} using GLSA and RA indicate that the TSB base flow is globally stable but amplifies low-frequency ($St \simeq 0.01$) and low-wavenumber ($\beta \simeq 1$) upstream perturbations. In effect, the flow behaves as a selective amplifier and the optimal energy gain computed from RA has the form of a first-order low-pass filter with a cut-off frequency of the order of $St \simeq 0.01$. In the present section we review these results and discuss them in light of the experimental database presented in section \ref{sec:expdata}.

\begin{figure}
\centering
    \begin{minipage}{\textwidth}
        \includegraphics[trim={5cm 9.6cm 5cm 3cm},clip,width=\textwidth]{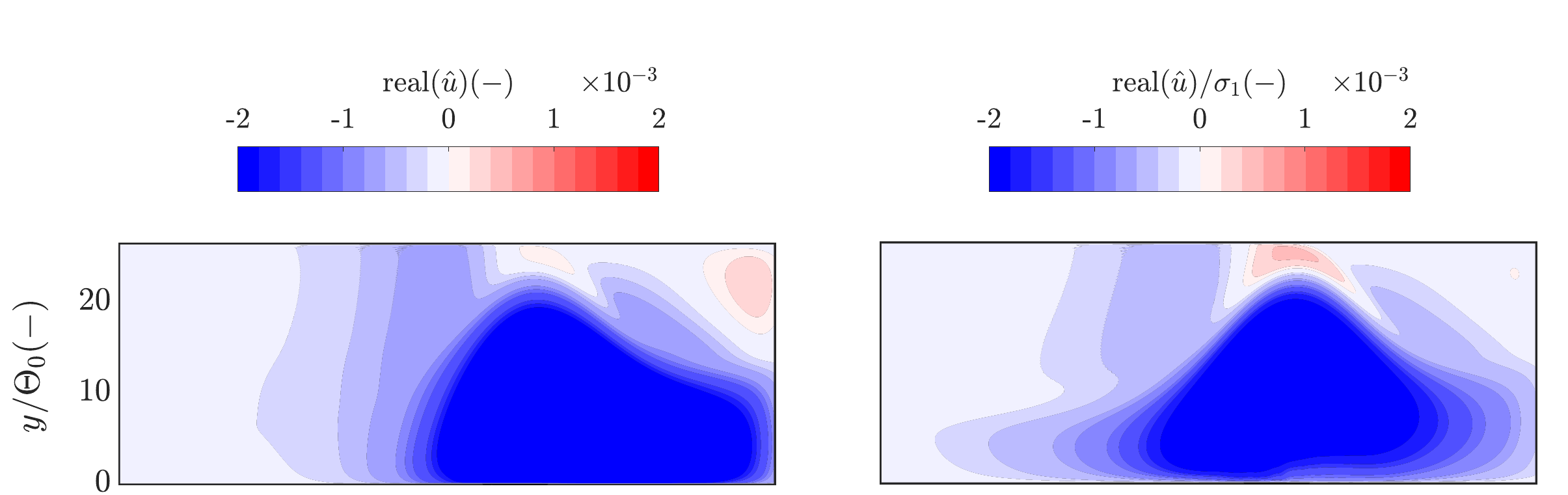}
            \put(-370,65){($a$)}
            \put(-185,65){($b$)}
    \end{minipage}
  \begin{minipage}{\textwidth}
   \includegraphics[trim={5cm 9.6cm 5cm 4.5cm},clip,width=\textwidth]{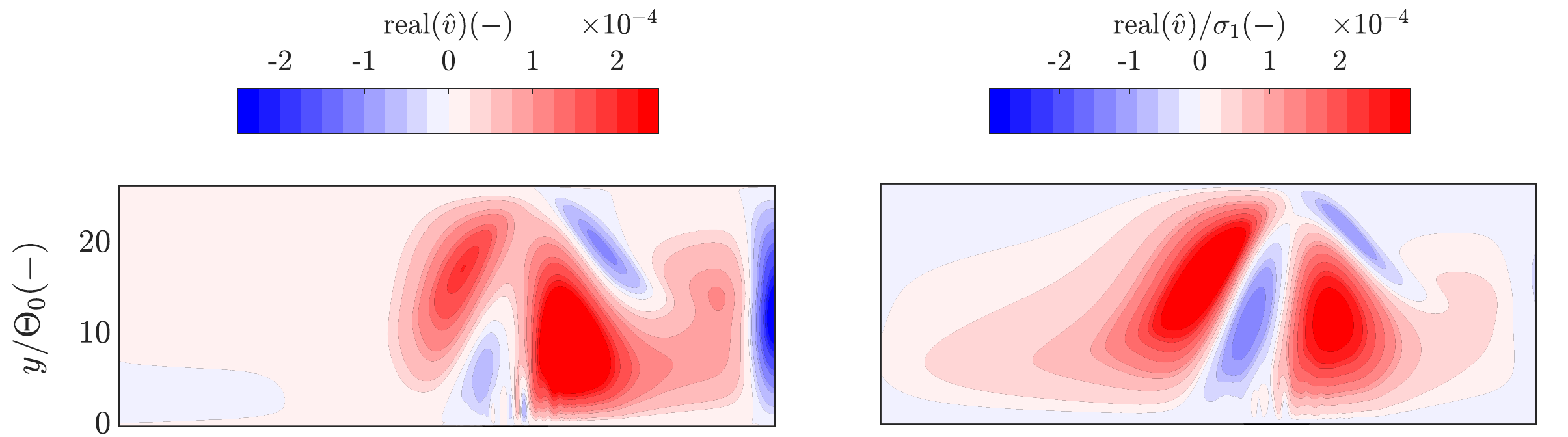}
                \put(-370,65){($c$)}
            \put(-185,65){($d$)}
   \end{minipage}
            \begin{minipage}{\textwidth}
       \includegraphics[trim={5cm 7.5cm 5cm 4.5cm},clip,width=\textwidth]{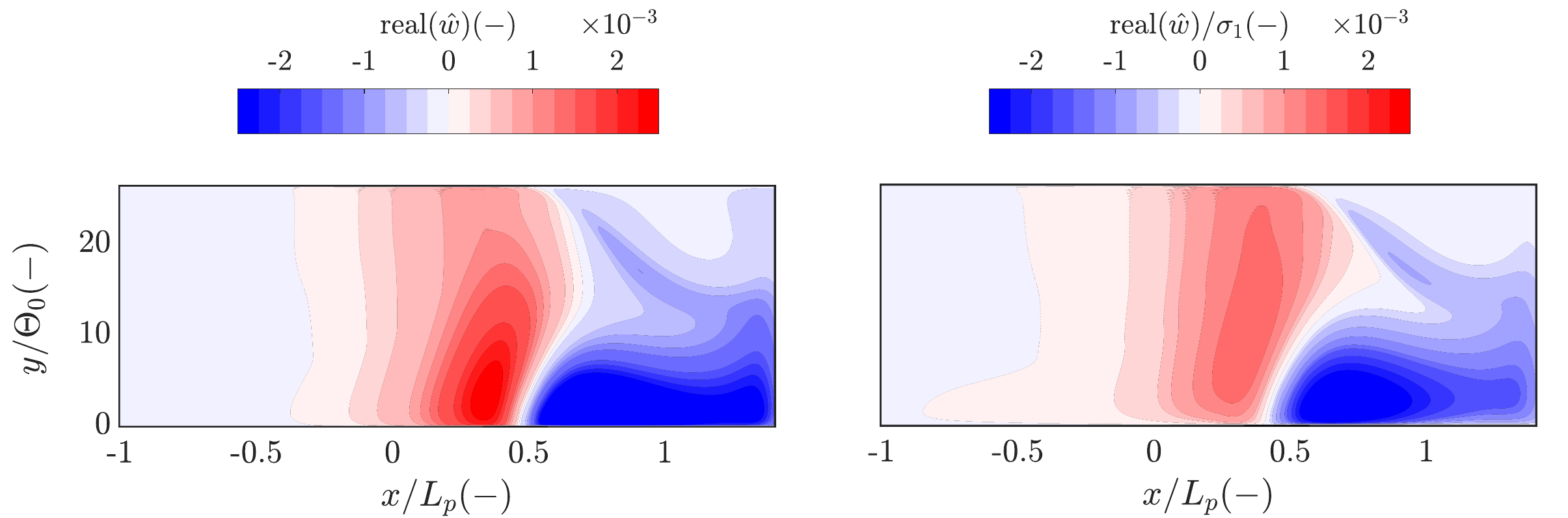}
                    \put(-370,85){($e$)}
            \put(-185,85){($f$)}
    \end{minipage}
\caption{Global LSA mode 1 ($a,c,e$) and optimal response from RA ($b,d,f$) for non-dimensional spanwise wavenumber $\beta=0.75$. The streamwise ($a,b$), wall-normal ($c,d$) and spanwise ($e,f$) modes are depicted, respectively.}
\label{fig:GLSAvsRA}
\end{figure}

We commence by comparing the least stable mode from GLSA (Mode 1) to the optimal response from RA at the fixed frequency $St=0.01$ and the non-dimensional spanwise wavenumber $\beta=0.75$. Once again, it is crucial to mention that no distinct difference could be observed in the low-frequency, low non-zero-$\beta$ regime in terms of the qualitative appearance of the RA modes. The streamwise ($a,b$), wall-normal ($c,d$), and spanwise component ($e,f$) of the respective modes are shown in figure  \ref{fig:GLSAvsRA}. The GLSA mode is shown on the left ($a,c,e$) and the optimal response from RA is shown on the right ($b,d,f$). The optimal response is normalized by the optimal gain $\sigma_1$ to ensure comparability.  It is apparent that all components $\hat u, \hat v, \hat w$ share strong similarities, with a comparable streamwise and wall-normal extent of the structures, a similar streamwise position $x/L_p$, and a matching phase. In the case of the streamwise modes $\hat u$ this equates to a large-scale structure bounding the recirculation region and following its shape, as described in the previous sections. Moreover, the amplitudes of the modes are in good agreement and the maximum is reached in a similar region of $x/L_p$. Similar trends can be observed for the wall-normal and spanwise modes. Here, the structures are almost a perfect match in terms of position, phase and size. The obvious similarities between GLSA and RA modes strongly suggest that the receptivity at low frequency is most likely occurring from the excitation of the least stable global mode of the TSB (Mode 1). 
\begin{figure}
\centering
    \begin{minipage}{0.49\linewidth}
        \includegraphics[trim={0 0 0 0},clip,width=0.95\textwidth]{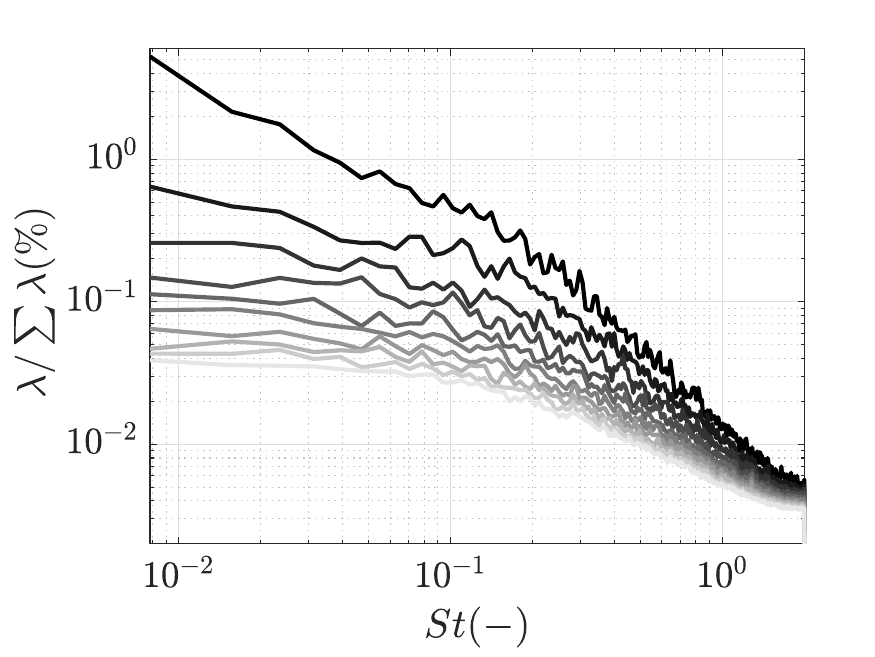}
          \put(-165,130){($a$)}
    \end{minipage}
    \begin{minipage}{0.49\linewidth}
        \centering
         \includegraphics[trim={0 0 0 0},clip,width=0.95\textwidth]{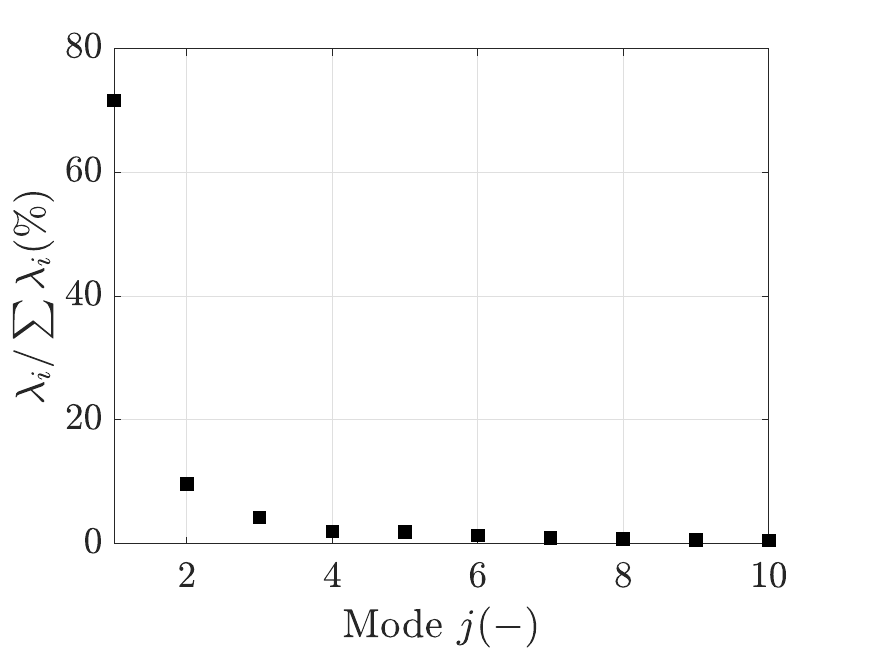}
           \put(-180,130){($b$)}
    \end{minipage}
\caption{Spectral POD eigenvalues $\lambda$ represented as the percentage of all modes ($a$) and SPOD eigenvalues $\lambda_i$ at Strouhal number $St=0.01$ represented as percentage of modes at low frequency ($b$). The modes $1-10$ are depicted, respectively.}
\label{fig:SPOD_Eigs}
\end{figure}

We now proceed to investigate the low-frequency unsteadiness observed in the experimental data in more details. Specifically, our objective is to establish a connection between the low-frequency unsteadiness (breathing) and the described low-frequency receptivity. For this purpose, we first demonstrate that the observed low-frequency dynamics in the experimental data capture a large portion of the total turbulent kinetic energy (TKE) of the flow, that is, that the SPOD eigenvalues are of \say{low rank} in the low-frequency regime. Then, we proceed to show analogies between RA and SPOD results. 

\begin{figure}
\centering
   \begin{minipage}{0.49\linewidth}
    \includegraphics[trim={0 0 0 0},clip,width=0.95\textwidth]{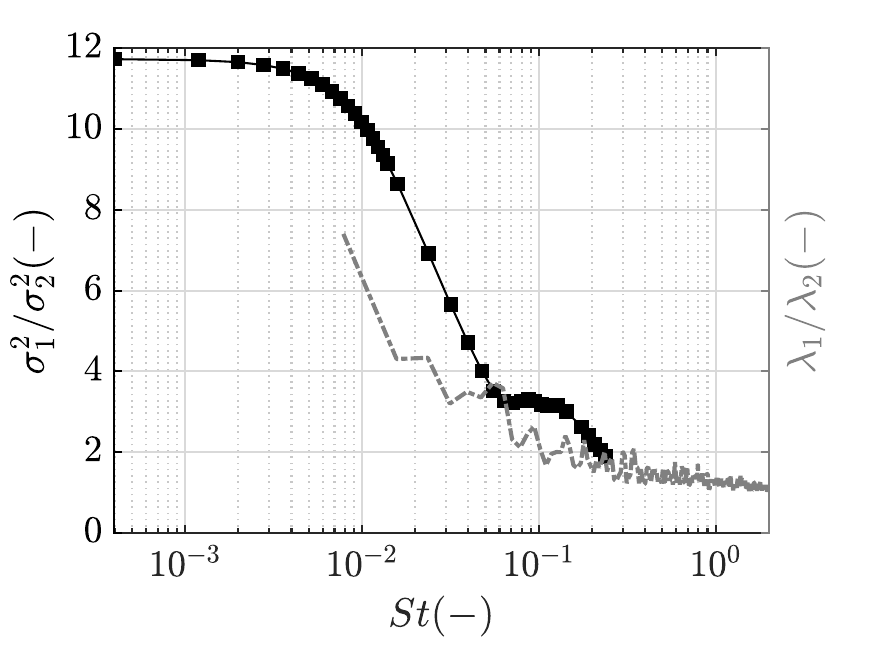}
   \end{minipage}    
\caption{Low-rank behavior of energy gains $\sigma_1^2/\sigma_2^2$ of resolvent analysis (\solidlinesquare) and SPOD eigenvalues $\lambda_1/\lambda_2$ (\dashedline) over different Strouhal numbers.}
\label{fig:lowrank}
\end{figure}

In figure \ref{fig:SPOD_Eigs} ($a$) the SPOD eigenvalues $\lambda$ are depicted for all modes and all Strouhal numbers $St$. The eigenvalues are represented as the percentage of the total turbulent kinetic energy (TKE) density, ensuring that the summation of all eigenvalues equals $\SI{100}{\percent}$. From figure \ref{fig:SPOD_Eigs} it becomes apparent that the first mode is low-rank in the low-frequency region, particularly for $St\simeq 0.01$. Therefore, if we investigate the first SPOD mode $\lambda_1$ at the fixed frequency $St=0.01$ (figure \ref{fig:SPOD_Eigs} ($b$)), we expect to capture a dominant feature of the flow. Here, the percentage of the TKE of the first mode is approximately $\SI{72}{\percent}$. Further evidence of the low-rank behavior of both the RA and SPOD modes is provided in figure \ref{fig:lowrank}. Here, we plot the ratio between the energy gains $\sigma_1^2/\sigma_2^2$ and between the SPOD eigenvalues $\lambda_1/\lambda_2$. In both cases, in the low-frequency range, the low-rank behavior is of the order of $10$.

\begin{figure}
\centering
\includegraphics[trim={0 0 0 0},clip,width=\textwidth]{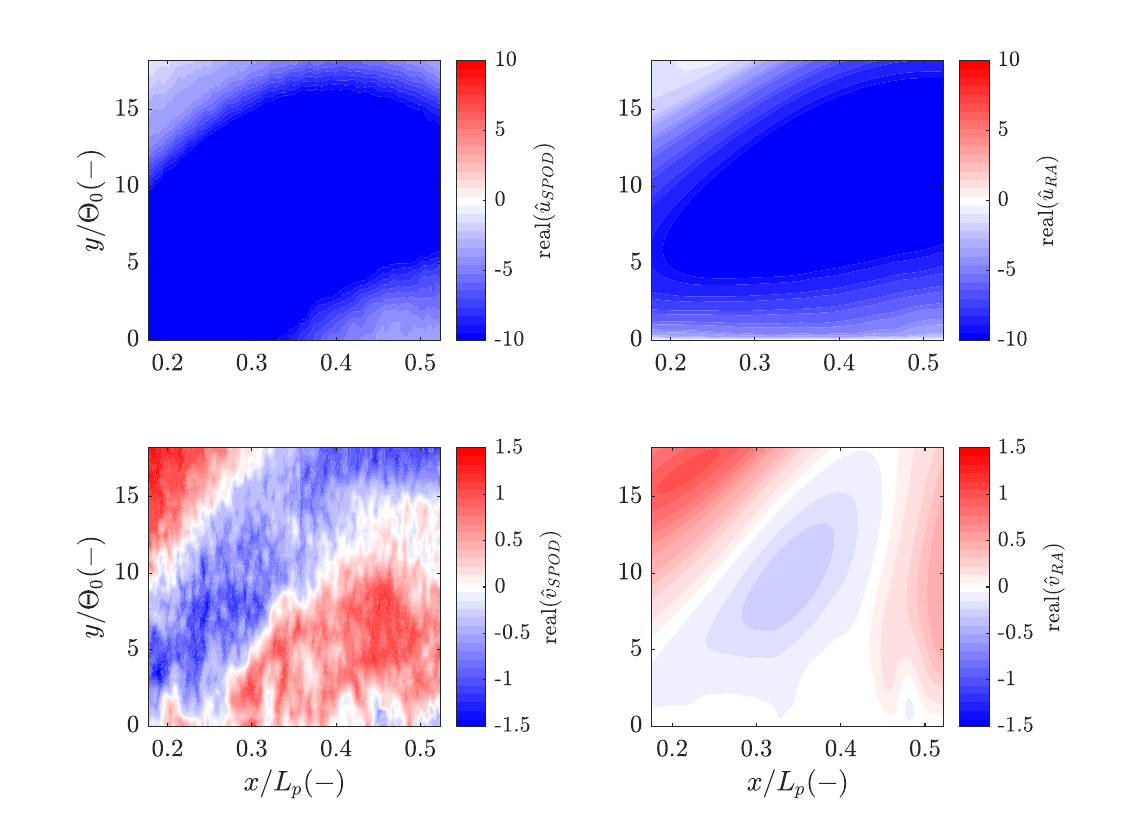}
            \put(-355,250){($a$)}
            \put(-185,250){($b$)}
            \put(-355,120){($c$)}
            \put(-185,120){($d$)}
\caption{Comparison between leading SPOD mode of the fluctuating velocity field from experiments ($a,c$) and optimal response of RA ($b,d$) in the region of the PIV measurements. Results at low frequency $St=0.01$ and with the highest alignment $\varphi=0.95$, $\beta=0.25$ are depicted. The streamwise ($a,b$) and wall-normal ($c,d$) component are displayed, respectively. All modes are scaled such that the wall-normal component is $|\hat v|=1$ and the phase is zero at a fixed position in space ($x/L_p=0.2, y/\Theta_0=16$).}
\label{fig:SPOD_alignment}
\end{figure}
\begin{figure}
\centering
\includegraphics[trim={0 0 0 0},clip,width=0.85\textwidth]{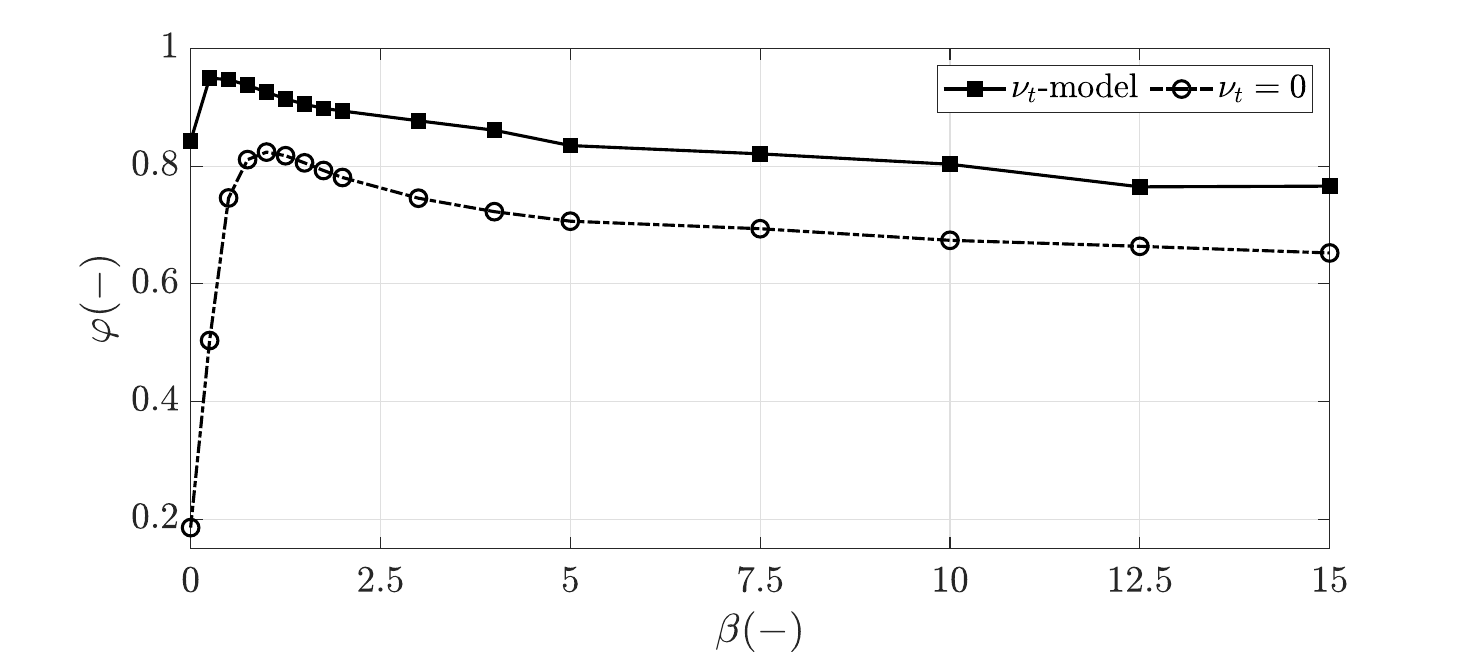}
\caption{Alignment $\varphi$ for different non-dimensional spanwise wavenumbers $\beta$. Results with (\solidlinesquare) and without (\dashedlinecircle) an eddy-viscosity model are shown.}
\label{fig:phi_beta}
\end{figure}
The comparison between the highest ranking SPOD mode ($a,c$) and the optimal response of RA ($b,d$) is shown in figure \ref{fig:SPOD_alignment} for $St=0.01$ and $\beta=0.25$. Here, only the $x/L_p$-region of the PIV measurement plane is depicted. Furthermore, we display the streamwise $\hat u$ ($a,b$) and wall-normal $\hat v$ ($c,d$) components, as only those are available from the experiment. In order to address the inherent arbitrariness of the phase in both SPOD and RA, the modes are subjected to a normalization. For this purpose, the phase of $\hat v$ is set to zero, while $|\hat v|=1$ at the fixed position $x/L_p=0.2$ and $y/\Theta_0=17$. From figure \ref{fig:SPOD_alignment} it becomes apparent that the first SPOD mode and the optimal gain of RA are in close agreement when $St=0.01$ and $\beta=0.25$. The streamwise component $\hat u$ displays a portion of the large structure surrounding the separation bubble as, e.g., previously shown in figure \ref{fig:RA_breathing}. The structures are tilted away from the solid bottom wall and follow the shape of the separated region. Furthermore, an excellent agreement between SPOD and RA can be observed in terms of amplitude and phase. The wall-normal component is also in good agreement with respect to the phase of the structures. A phase shift oriented approximately perpendicular to the shear layer can be observed. The amplitude, however, is slightly larger in the case of the first SPOD mode. The alignment $\varphi$, as introduced in section \ref{sec:methods}, is $\varphi=0.95$. This large value of $\varphi$ strongly suggests that the low-frequency breathing observed in the experiment is associated with the calculated RA response.

We have previously established that any RA mode with low non-zero $\beta$ exhibits the same low-frequency characteristics when $St\approx 0.01$. In figure \ref{fig:phi_beta} we display the alignment $\varphi$ between SPOD and RA over the non-dimensional spanwise wavenumber $\beta$ at the fixed frequency $St=0.01$. Both the analysis including an eddy-viscosity model and the analysis where $\nu_t=0$, are depicted. We obtain two distributions of comparable qualitative nature. However, a strong offset between the alignment for the analysis including an eddy-viscosity model and the analysis where $\nu_t=0$ is evident. Clearly, including a $\nu_t$-model into the LNSE operator strongly increases the alignment with the experiment, as represented by the SPOD modes. This is particularly relevant since the overall shape of the modes is the same whether or not a $\nu_t$-model is included. Interestingly, the significance of implementing an eddy-viscosity model to enhance the alignment between SPOD and RA has been demonstrated in prior studies. For example, \cite{morra2019relevance} presented evidence of its relevance in turbulent wall-bounded flows, while \cite{pickering2020lift} extended this insight to a turbulent jet configuration. 

In summary, the results presented so far demonstrate a strong alignment between the SPOD of the streamwise/wall normal velocity fluctuations obtained experimentally at low frequency and the optimal RA response computed in the low non-zero-$\beta$ region. We further note that both the near wall velocity and wall-pressure fluctuations in the experiments indicate a characteristic wavenumber of the order of $\beta \simeq 1$ for the low-frequency breathing motion (e.g., figure \ref{pic:Rpp}). This cements the argument that the unsteadiness observed in the experiment is well described by the RA results. Hence, our conclusion is that the low-frequency breathing motion likely emanates from the excitation of the most weakly damped global mode in the TSB. This phenomenon is not dictated by a singular frequency but rather encompasses a range of low frequencies and low, non-zero spanwise wavenumbers. In effect, the TSB acts like a first-order low pass filter for upstream disturbances in the incoming turbulent boundary layer.

Interestingly, a low-pass filter model of low-frequency unsteadiness for separation bubbles has already been suggested in different flow configurations. \cite{mohammed2021}, in a  configuration similar to the present one, used upstream forcing to investigate the response of their TSB to upstream perturbations. They noted that the transient response of the separation line was compatible with a first-order low-pass filter model and observed consistent low-frequency dynamics in the unforced flow. For the case of turbulent SBLIs, \cite{poggie2015} investigated the frequency spectra of wall-pressure fluctuations in a variety of test cases and showed that despite differences in Reynolds and Mach numbers, the first-order model originally proposed by \cite{Plotkin1975} and refined by \cite{Touber2011} for shock-induced separation reasonably collapsed the results. Recently, \cite{bugeat2022low} proposed a similar low-pass-filter model for a laminar SBLI and demonstrated that this behaviour proceeds \say{from the excitation of a single, stable, steady global mode whose damping rate sets the time scale of the filter}. Hence, in the case of our TSB, the low-pass filter model is a direct consequence of our proposed mechanism of low-frequency unsteadiness since it is also based on the excitation of a stable global mode. Given the prevalence of low-pass filter behaviour in the literature on separation bubble unsteadiness, these results, combined with those of \cite{bugeat2022low} suggest that such a mechanism of low-frequency unsteadiness might occur in many types of separation bubbles.

Finally, we briefly come back to our original strategy of performing linear analysis on a DNS base flow that is similar, but not exactly the same, as the experimental flow field that we consider. In our case, the cross-validation between experimental unsteady data and the DNS base flow was necessary since the DNS did not include any unsteady data at the proper frequency, and the experimental data was not sufficiently resolved spatially to enable a proper linear analysis. Nevertheless, the concordance between the GLSA/RA results and the experimental data turned out to be close. This suggests that the phenomenology discovered in this work is not very sensitive to the exact geometry of the flow. Furthermore, the fact that the DNS base flow is devoid of any experimental artefacts like inflow unsteadiness, roughness effects, or the presence of wind-tunnel side walls, confirms that the low-frequency breathing motion observed in several low-speed and many high-speed experiments is a relevant fluid-dynamical phenomenon that is unlikely to be caused by wind-tunnel effects. 
\section{Conclusion}
\label{sec:concl}
The objective of the present work was to perform modal and non-modal stability analysis on an incompressible, pressure-gradient-induced TSB to investigate the origin of the low-frequency breathing motion commonly observed in experimental studies. Specifically, we performed global linear stability analysis (GLSA) and resolvent analysis (RA) on a base flow consisting of the average flow field computed via DNS by \cite{coleman2018}, and we compared the results with the unsteady experimental measurements obtained by \cite{le2018spanwise,le2020} in a similar flow field.

GLSA revealed that the investigated TSB is globally stable in the asymptotic time limit for all frequencies and wavenumbers, indicating that no self-sustained (intrinsic) instability of the flow is driving the low-frequency unsteadiness. Interestingly, a region of low non-zero $\beta$ was detected, for which the growth rate $\omega_i$ was closer to the stability threshold. This region was found to agree very well with the range of spanwise wavenumbers extracted from the experimental database at low frequency.

 At low-frequency and low non-zero spanwise wavenumber, RA further revealed an optimal response strongly resembling the unsteady velocity signature of the breathing motion, which typically manifests as a large-scale coherent structure that follows the shape of the bubble. Specifically, a comparison between the first SPOD mode of the streamwise and wall-normal velocity fluctuations measured experimentally showed strong alignment with the optimal RA response. This, combined with the stable GLSA modes, indicates that the TSB behaves as an amplifier flow in the low-frequency regime. Furthermore, the optimal gain computed by RA at low non-zero wavenumber was shown to have the characteristic shape of a first-order, low-pass filter with a cut-off frequency consistent with the low-frequency unsteadiness observed experimentally. 

 Finally, comparison between the least stable GLSA mode and the optimal RA response showed strong similarities in the low frequency and low wavenumber regime, and this for all velocity components. This behavior is considered a strong indication that the low-frequency oscillation is primarily driven by the least stable global mode, thereby characterizing the phenomenon as a forced modal mechanism, associated here with a centrifugal mechanism typical of other separation bubbles \citep{theofilis2000, rodriguez2013two}. Hence, the main conclusion of this study is that the low-frequency breathing of the TSB is most likely driven by the excitation of a weakly damped global mode by the turbulent fluctuations in the incoming boundary layer. This is consistent with the mechanism recently proposed by \cite{porter2019} for the case of turbulent SBLI flows. Our results also bears strong resemblance to the analysis of \cite{bugeat2022low} in a laminar SBLI, thereby suggesting that forced dynamics might be a common feature of separation bubbles in a wide range of flow conditions.
\appendix
\section{Matrix Operators}\label{appA}
In this section, the matrix operators employed in the global stability and resolvent analysis are described. We commence by revisiting the equation
\begin{equation}
    -i\omega\textbf{M} \hat{\textbf{q}}=(\textbf{A}+\textbf{N})\hat{\textbf{q}}+\textbf{B}\hat{\textbf{f}}.
\end{equation}
The linearized Navier-Stokes equations can be recast into an operator of the form 
\begin{equation}
 \textbf{A}=\left[\begin{array}{cccc}
\xi+\frac{\partial \overline{u}}{\partial x}-\frac{1}{Re}\nabla^2 & \frac{\partial \overline{u}}{\partial y} & 0 & \frac{\partial}{\partial x}	\\
\frac{\partial \overline{v}}{\partial x}	 & \xi+\frac{\partial \overline{v}}{\partial y}-\frac{1}{Re}\nabla^2 & 0 & \frac{\partial}{\partial y}	\\
0 & 0 & \xi-\frac{1}{Re}\nabla^2 & i\beta  	\\
\frac{\partial}{\partial x} & \frac{\partial}{\partial y} & i\beta & 0
\end{array}\right],
\end{equation}  
where only the molecular viscosity in the flow is considered. Here $\xi=(\overline{u}\frac{\partial}{\partial x}+\overline{v}\frac{\partial}{\partial y})$ and $\nabla^2=(\frac{\partial^2}{\partial x}+\frac{\partial^2}{\partial y}-\beta^2)$ is the Laplacian.\\

When the turbulence of the flow is considered by means of an eddy-viscosity model, the additional terms read
\begin{equation}
 \textbf{N}=\left[\begin{array}{cccc}
-\eta-\frac{\partial \nu_{t}}{\partial x} \frac{\partial}{\partial x} -\nu_{t}\nabla^2 & -\frac{\partial \nu_{t}}{\partial y} \frac{\partial}{\partial x}& 0 & 0 \\
-\frac{\partial \nu_{t}}{\partial x}\frac{\partial}{\partial y}  & -\eta-\frac{\partial \nu_{t}}{\partial y} \frac{\partial }{\partial y}-\nu_{t}\nabla^2& 0 & 0\\
 -i\beta \frac{\partial \nu_{t}}{\partial x}  & -i\beta \frac{\partial \nu_{t}}{\partial y} & -\eta-\nu_{t}\nabla^2 & 0\\
  0  & 0 & 0 & 0\\ 
\end{array}\right],
\end{equation}
where $\eta=(\frac{\partial \nu_{t}}{\partial x}\frac{\partial}{\partial x}+\frac{\partial \nu_{t}}{\partial y}\frac{\partial}{\partial y})$.
The operator $\textbf{M}$ acts like a mass-like matrix in order to recover the continuity equation
\begin{equation}
\textbf{M}=\left[\begin{array}{cccc}
   \mathcal{I}  & 0 & 0 & 0\\ 
  0  &  \mathcal{I} & 0 & 0\\ 
   0   & 0 &  \mathcal{I} & 0\\ 
   0 & 0 & 0 & 0 \\ 
\end{array}\right],
\end{equation}
where  $ \mathcal{I}$ is the identity matrix of respective size. In the resolvent form
\begin{equation}
  \hat{\textbf{q}}=\textbf{C}(-i \omega \textbf{M}-\textbf{A}_{2D,z})^{-1}\textbf{B}\hat{\textbf{f}},  
\end{equation}
we define the additional matrix operators $\textbf{B}$ and $\textbf{C}$ 
\begin{equation}
\textbf{B}=\left[\begin{array}{ccc}
   b(x)  & 0 & 0 \\ 
  0  &  b(x) & 0 \\ 
   0   & 0 &  b(x) \\ 
   0 & 0 & 0 \\ 
\end{array}\right],
\hspace{0.5cm}
\textbf{C}=\left[\begin{array}{cccc}
   c(x)  & 0 & 0 & 0\\ 
  0  &  c(x) & 0 & 0\\ 
   0   & 0 &  c(x) & 0\\ 
\end{array}\right].
\end{equation}
The sparse diagonal matrices act as filters that enforce constraints on the forcing and response, respectively. Here, $b(x)=1$ in the region $x\in[0.05L_x, 0.85L_x]$, whereas $b(x)=0$ in the remainder of the domain. Similarly, we set $c(x)=1$ when $x\in[0.33L_x,0.74L_x]$ and $c(x)=0$ elsewhere. $L_x$ is the length of the domain. The region $x\in[0.33L_x,0.74L_x]$ is chosen as to agree with the PIV measurement region, which only covers a portion of the domain of the numerical data.
\section{Boundary Conditions and Fringe Zone}\label{appB}
\begin{figure}
    \centering
\includegraphics[trim={0 0 0 0},clip,width=0.6\textwidth]{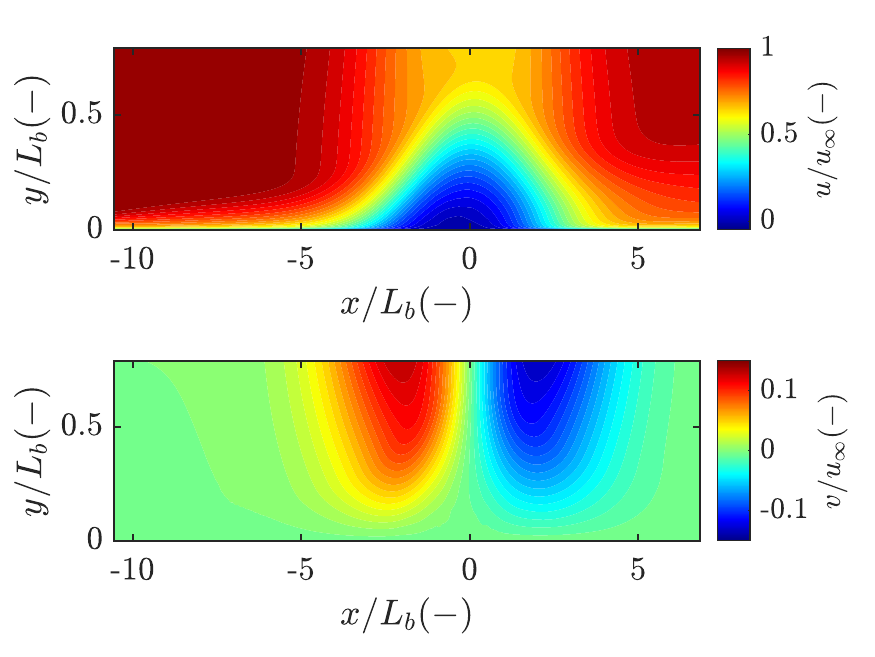}
 \put(-220,165){($a$)}
  \put(-220,83){($b$)}
    \caption{Non-dimensional DNS base flow using the length scale $l^*=L_b$ and the time scale $t^*=L_b/u_\infty$. The streamwise $u/u_\infty$ ($a$) and wall-normal $v/u_\infty$ ($b$) component of the time-averaged velocity field are displayed.}
    \label{fig:DNSuv}
\end{figure}

The non-dimensional DNS base flow from \cite{coleman2018}, using the length scale $l^*=L_b$ and the time scale $t^*=L_b/u_\infty$, is depicted in Fig. \ref{fig:DNSuv}. In the global stability and resolvent analysis, we subject the DNS base flow to periodic boundary conditions along the streamwise direction
\begin{equation}
    u(x+L_\pi)=u(x),
    \end{equation}
    \begin{equation}
    v(x+L_\pi)=v(x),
\end{equation}
where the periodicity length $L_\pi$ is chosen according to $L_\pi=0.9N_x$ and $N_x$ is the streamwise dimension of the velocity fields. We further enforce that the disturbances decay at the solid top ($\infty$) and bottom wall ($0$) according to
\begin{equation}
    \hat u(0)=\hat v(0)=\hat w(0)=0,
\end{equation}
\begin{equation}
        \hat u(\infty)=\hat v(\infty)=\hat w(\infty)=0.
\end{equation}

In order to suppress all disturbances $\hat u, \hat v, \hat w$ as well as forcings $\hat f_x, \hat f_y, \hat f_z$ in the region where the periodic boundary conditions are enforced, a fringe 
\begin{equation}
\textbf{S}=\left[\begin{array}{cccc}
   \sigma_f  & 0 & 0 & 0\\ 
  0  &  \sigma_f & 0 & 0\\ 
   0   & 0 &  \sigma_f & 0\\ 
   0 & 0 & 0 & 0 \\ 
\end{array}\right],
\end{equation}
is added to the two-dimensional LNSE operator $\textbf{A}_{2D,z}=\textbf{A}+\textbf{N}+\textbf{S}$. Here, $\sigma_f$ is a sparse diagonal matrix with $\sigma_f = s_\textrm{max}\cdot \textrm{exp}(-(x-x_c)^2/L_f^2)$, $s_\textrm{max}$ is the maximum amplitude of the fringe, $x_c$ is the center of the fringe and $L_f$ is the length. The position and length of the fringe are chosen as $x_c\approx 6L_b$ and $L_f\approx\SI{3.5}{\percent}L_x$, where $L_x$ is the length of the domain.
\section{Grid and Fringe Convergence}\label{appC}
\begin{figure}
\centering
\includegraphics[trim={0 0 0 0},clip,width=\textwidth]{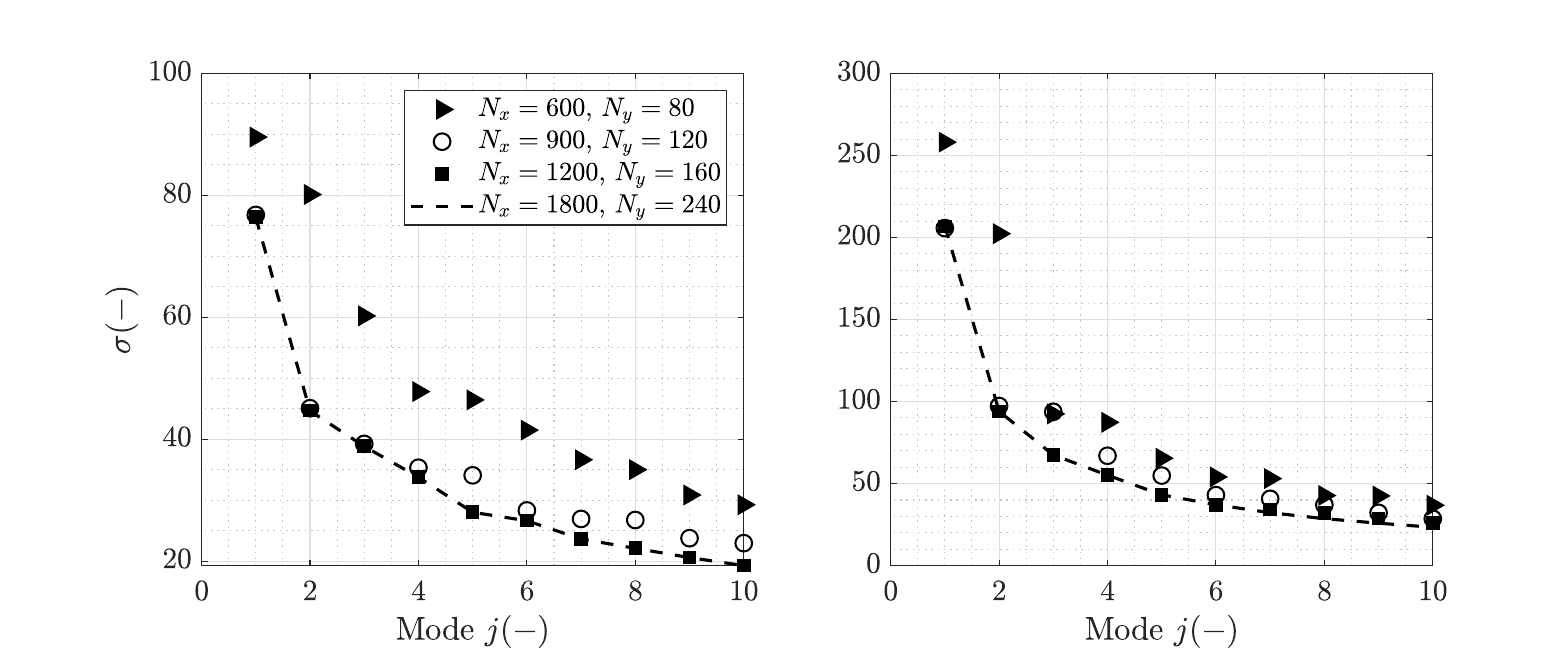}
 \put(-360,150){($a$)}
  \put(-190,150){($b$)}
\caption{Grid convergence study for $St=0.01$ ($a$) and $St=0.1$ ($b$) based on Resolvent Analysis at $\beta=0$. Grid independence of the singular values of the RA can be observed for $N_x\ge 1200$ and $N_y \ge 160$.}
\label{fig:GridConvergence}
\end{figure}

The grid convergence study is presented in Fig. \ref{fig:GridConvergence}. We portray the first ten singular values of the resolvent analysis at $\beta=0$. Two Strouhal numbers are displayed: a Strouhal number $St=0.01$, representative of the characteristic low-frequency breathing motion ($a$), and a  Strouhal number in the medium-frequency regime $St=0.1$ ($b$). Four different streamwise ($N_x$) and wall-normal ($N_y$) resolutions are depicted. For a resolution of $N_x$ x $N_y$ equal to $1200$ x $160$ or higher, a convergence of the results can be observed. Further increasing the grid resolution yields no discernible influence on the singular values of the RA.

\begin{figure}
\centering
\includegraphics[trim={0 0 0 0},clip,width=\textwidth]{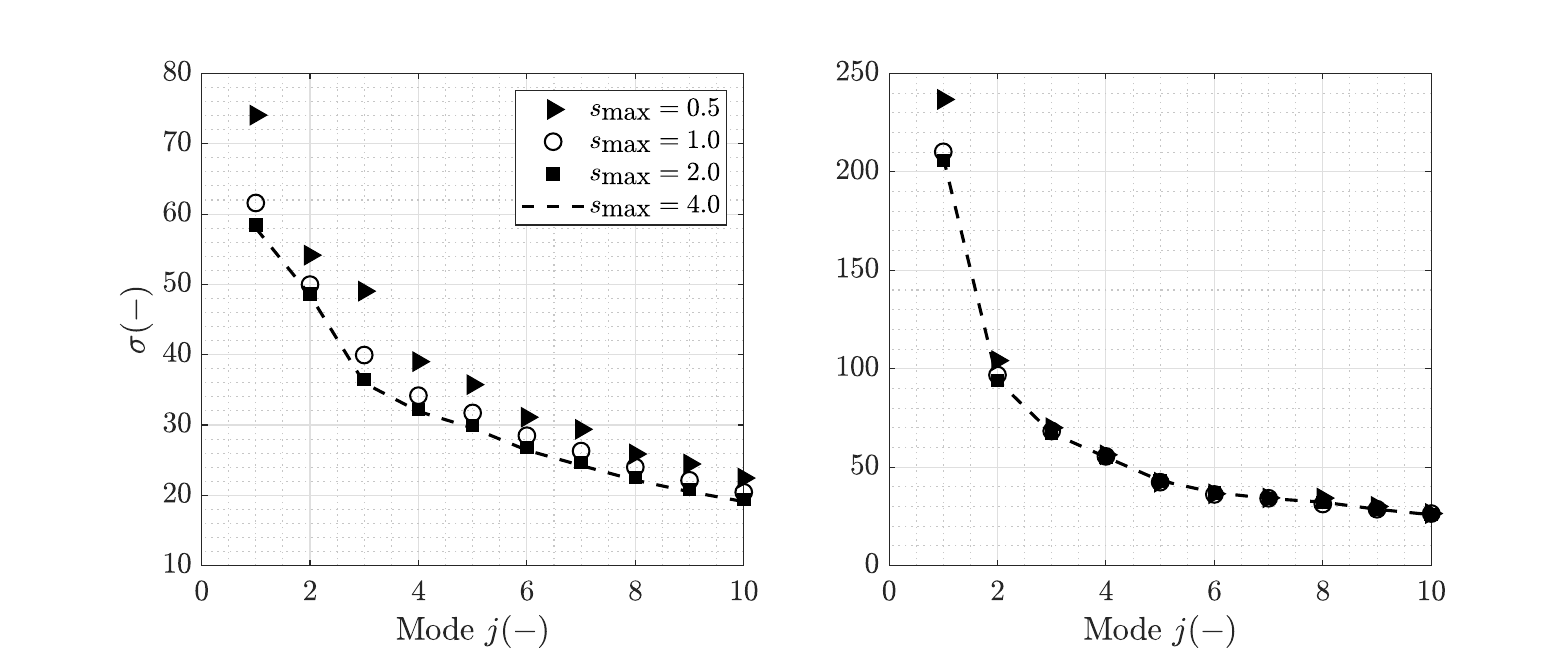}
 \put(-360,150){($a$)}
  \put(-190,150){($b$)}
\caption{Fringe convergence study for $St=0.01$ ($a$) and $St=0.1$ ($b$) based on resolvent analysis at $\beta=0$. Fringe independence of the singular values of the RA can be observed for $s_\textrm{max}\ge2$.}
\label{fig:FringeConvergence}
\end{figure}

We illustrate the fringe convergence study in Fig. \ref{fig:FringeConvergence}. Once again, we display $St=0.01$ ($a$) and $St=0.1$ ($b$). We maintain a fixed resolution $N_x=1200$ and $N_y=160$ for the analysis and test four different magnitudes $s_\textrm{max}$. Fringe convergence can then be observed for magnitudes $s_\textrm{max}\ge 2$.\\
In this work we perform the GLSA and RA analyses with a fringe magnitude of $s_\textrm{max}=5$. All resolvent analyses are performed with a  resolution of $N_x=1200$ and $N_y=160$. A slightly higher resolution of $N_x=1200$ and $N_y=320$ is employed in the case of GLSA. 
\newpage
\bibliographystyle{style}
\bibliography{literature}

\end{document}